\documentclass[sigconf]{acmart}

\acmConference[MSR 2022]{MSR '22: Proceedings of the 19th International
Conference on Mining Software Repositories}{May 23–24, 2022}{Pittsburgh,
PA, USA}

\usepackage{color}
\usepackage{xcolor}
\usepackage{multirow}
\usepackage{ifthen}
\usepackage{url}
\usepackage{amsmath}

\usepackage{hyphenat}
\usepackage{amssymb}
\usepackage{xspace}
\usepackage{enumitem}
\usepackage[ruled,linesnumbered]{algorithm2e}
\usepackage{array,multirow,graphicx}
\usepackage{tikz}
\usepackage{calc}
\usepackage[caption=false]{subfig}
\usepackage{listings,amsfonts}
\usepackage{amsmath,pifont}
\usepackage{epsfig}
\usepackage{endnotes}
\usepackage{makecell}
\usepackage[normalem]{ulem}
\usepackage{pgfplots}
\usepackage{capt-of}
\usepackage{tcolorbox}
\usepackage{adjustbox}
\usepackage{amsthm}
\usepackage{lipsum}
\usepackage{balance}
\usepackage{todonotes}

\newcommand{\cmmnt}[1]{}

\newcommand{\ie}{i.e.,\xspace}
\newcommand{\eg}{e.g.,\xspace}
\newcommand{\etc}{etc.\xspace}

\pgfplotsset{width=7cm,compat=1.16}
\usepgfplotslibrary{statistics}
\graphicspath{ {./} }

\definecolor{yellow}{RGB}{255,255,153}
\definecolor{grey}{RGB}{224,224,224}
\definecolor{green}{RGB}{0,100,0}

\newboolean{showcomments}
\setboolean{showcomments}{true}
\ifthenelse{\boolean{showcomments}}
 { \newcommand{\mynote}[2]{
      \fbox{\bfseries\sffamily\scriptsize#1}
        {\small$\blacktriangleright$\textsf{\emph{#2}}$\blacktriangleleft$}}}
        { \newcommand{\mynote}[2]{}}

\definecolor{DarkOrange}{rgb}{0.8,0.3,0.0} 
\definecolor{DarkCyan}{rgb}{0.0, 0.55, 0.55}
\definecolor{codegreen}{rgb}{0,0.6,0}
\definecolor{codegray}{rgb}{0.5,0.5,0.5}
\definecolor{codepurple}{rgb}{0.58,0,0.82}
\definecolor{backcolour}{rgb}{0.95,0.95,0.92}

\newcommand{\PP}[1]{
\vspace{3px}
\noindent{\bf \textsc{\IfEndWith{#1}{.}{#1}{#1.}}}
}

\AtBeginDocument{%
  \providecommand\BibTeX{{%
    \normalfont B\kern-0.5em{\scshape i\kern-0.25em b}\kern-0.8em\TeX}}}

\copyrightyear{2022}
\acmYear{2022}
\setcopyright{acmcopyright}\acmConference[MSR '22]{19th International Conference
on Mining Software Repositories}{May 23--24, 2022}{Pittsburgh, PA, USA}
\acmBooktitle{19th International Conference on Mining Software Repositories (MSR
'22), May 23--24, 2022, Pittsburgh, PA, USA}
\acmPrice{15.00}
\acmDOI{10.1145/3524842.3527963}
\acmISBN{978-1-4503-9303-4/22/05}

\begin{document}

\title{Do Customized Android Frameworks\\Keep Pace with Android?}
%Do Customized Android Frameworks Keep Pace with Android?
%An In-depth Study of the Merge Operations Characterizing Android Framework Customizations
%An In-depth Study of Customized FrameworkEvolution in the Android Ecosystem
%How do Customized Android Frameworks Co-evolve with Official Android?

% \author{Pei Liu, Mattia Fazzini, John Grundy, and Li Li
% \IEEEcompsocitemizethanks{
% \IEEEcompsocthanksitem Pei Liu, John Grundy, and Li Li are with Monash University, Australia. Li Li is the corresponding author.
% \IEEEcompsocthanksitem Mattia Fazzini is with the University of Minnesota, United States. 
% }
% }

\author{Pei Liu}
\email{pei.liu@monash.edu}
\affiliation{
  \institution{Monash University}
  \city{Melbourne}
  \state{Victoria}
  \country{Australia}}

\author{Mattia Fazzini}
\email{mfazzini@umn.edu}
\affiliation{
  \institution{University of Minnesota}
  \city{Minneapolis}
  \state{Minnesota}
  \country{United States}}

\author{John Grundy}
\email{john.grundy@monash.edu}
\affiliation{
  \institution{Monash University}
  \city{Melbourne}
  \state{Victoria}
  \country{Australia}}

\author{Li Li}
\authornote{Corresponding author.}
\email{li.li@monash.edu}
\affiliation{
  \institution{Monash University}
  \city{Melbourne}
  \state{Victoria}
  \country{Australia}}

% \markboth{Journal of \LaTeX\ Class Files,~Vol.~xx, No.~xx, xx~2020}%
% {Shell \MakeLowercase{\textit{et al.}}: titleXXX}

% \IEEEtitleabstractindextext{%

\begin{abstract}
To satisfy varying customer needs, device vendors and OS providers often rely on the open-source nature of the Android OS and offer customized versions of the Android OS.
When a new version of the Android OS is released, device vendors and OS providers need to merge the changes from the Android OS into their customizations to account for its bug fixes, security patches, and new features. Because developers of customized OSs might have made changes to code locations that were also modified by the developers of the Android OS, the merge task can be characterized by conflicts, which can be time-consuming and error-prone to resolve.
%This characteristic and the limited support for the task might lead OS customizations to even skip some of the updates.

%
To provide more insight into this critical aspect of the Android ecosystem, we present an empirical study that investigates how eight open-source customizations of the Android OS merge the changes from the Android OS into their projects.
The study analyzes how often the developers from the customized OSs merge changes from the Android OS, how often the developers experience textual merge conflicts, and the characteristics of these conflicts. Furthermore, to analyze the effect of the conflicts, the study also analyzes how the conflicts can affect a randomly selected sample of 1,000 apps.
After analyzing 1,148 merge operations, we identified that developers perform these operations for 9.7\% of the released versions of the Android OS, developers will encounter at least one conflict in 41.3\% of the merge operations, 58.1\% of the conflicts required developers to change the customized OSs, and 64.4\% of the apps considered use at least one method affected by a conflict.
In addition to detailing our results, the paper also discusses the implications of our findings and provides insights for researchers and practitioners working with Android and its customizations.
\end{abstract}

% \begin{IEEEkeywords}
% Android, Framework Evolution, Customized Android Framework, Merge Conflicts
% \end{IEEEkeywords}

% }

\maketitle

\section{Introduction}
\label{sec:introduction}

Today, we use mobile devices for many of our daily activities, such as reading the news, doing online shopping, streaming content, and communicating with family and friends. One common trait of mobile devices is that they rely heavily on their operating system (OS). In fact, mobile devices use the OS to manage applications (apps), facilitate inter-app communication, and allow apps to access a device's hardware.
The Android OS is currently the most widely used OS across mobile devices, having 87\% of the market share~\cite{2021_statista} and running on more than two billion devices~\cite{2021_verge_android}. One reason behind Android's popularity is that the OS is an open platform. This characteristic enables device vendors and OS providers to satisfy a large variety of customer needs with customized OS versions.

To create a customized version of the Android OS, device vendors and OS providers modify their own copy of the source code of the Android OS, whose official version is available in the repository managed by the Android Open Source Project (AOSP)~\cite{2021_aosp}. When device vendors and OS providers create customized versions of the Android OS, they usually add new features but may also modify or even delete existing ones. These customizations often contain changes that significantly differ from the original code of the Android  OS, leading to divergent versions of the Android OS.
Even though these customizations are divergent versions of the Android OS, device vendors and OS providers periodically need to update their customizations so that they can integrate the changes from the Android OS, which provide bug fixes, security patches, and new functionalities~\cite{li2016accessing, li2020cda, gao2019understanding}. To this end, developers from customized OSs perform merge operations based on the new releases of the Android OS. Because such merge operations can span across the changes characterizing the customized OSs, these operations can, unfortunately, lead to merge conflicts.

While software developers have always experienced some form of merge conflicts in non-divergent team projects, the conflicts experienced in the customizations of the Android OS might be particularly challenging to be resolved. This situation can emerge because (i) the changes in the Android OS happen without any knowledge of the developers working on the customized OSs; (ii) the Android OS evolves at a very rapid pace~\cite{mcdonnell2013empirical, linares2013api, bavota2014impact, yang2018how, li2018moonlightbox, li2018characterising}; and (iii) new releases of the Android OS frequently contain thousands of commits.
%These characteristics and the limited support for the task might even lead OS customizations to neglect some of the Android updates.
Although related work has analyzed how some of the changes in one customization of the Android OS compare to some of the changes in the Android OS~\cite{mahmoudi2018android}, there is still little understanding on whether customized OSs merge the changes from new releases of the Android OS and the characteristics of the actual merge operations. Furthermore, we also do not yet know how different customizations of the Android OS perform this type of merge operation and whether merge conflicts are a recurring problem in different customizations.

To bridge this gap, we present an empirical study that analyzes the version control history of eight open-source customizations of the Android OS and investigates how their developers merged the changes from the new releases of the Android OS into their projects. The study focuses on the portion of the Android OS that provides the Android framework base (as this part contains key OS services that apps directly and heavily rely on~\cite{mcdonnell2013empirical, linares2013api, bavota2014impact, yang2018how, li2018moonlightbox, li2018characterising, 2019_issta_fazzini_automated}).
The study investigates how frequently developers performed this type of merge operations and the properties of the operations. Furthermore, to provide a perspective on the potential effects of these conflicts, we also analyzed a randomly selected sample of 1,000 apps and studied how many apps use methods that are affected by conflicts.
Our results show that (1) developers performed this type of merge operation for 9.7\% of versions released by the Android OS, (2) developers have encountered at least one conflict in 41.3\% of the merge operations performed, (3) source code methods are the code entities most affected by the conflicts, (4) 58.1\% of the conflicts required developers to change the customized OSs, and (5) 64.4\% of the considered apps use at least one method affected by a conflict.

The large number of conflicts identified and the low percentage of merge operations performed motivate further research aiming to help developers perform these operations. Furthermore, the high number of conflicts and the high percentage of apps using methods affected by conflicts indicate that these apps might also experience a range of compatibility issues (\ie issues that prevent apps from working as expected when running on customized OSs~\cite{wei2016taming, he2018understanding, li2018cid, scalabrino2019data, xia2020android, cai2019large}) and further research is needed to help app developers handle these issues. The paper elaborates on our findings to help researchers and practitioners who work with the Android OS and guide future research on the topic.

%\PP {Contributions and Significance.}
%
This paper presents an empirical study of how different customizations of the Android OS merge changes to account for new releases of the Android OS. The paper also
analyzes the extent to which the conflict-affected methods are accessed in Android apps.
Our findings and their implications can help in the design of automated or semi-automated techniques for better supporting developers of
customized Android OSs. Finally, to support future research, we make our study infrastructure and results publicly available in our online appendix~\cite{customizedscripts}.

\newsavebox\androidbefore
\begin{lrbox}{\androidbefore}
\hspace{-45pt}
\begin{lstlisting}[escapechar=|,language=Java,basicstyle=\scriptsize\ttfamily,numbers=left,numbersep=5pt, numberstyle=\scriptsize\color{black},frame=None,xleftmargin=15pt,xrightmargin=2pt, aboveskip=5pt,belowskip=5pt]
...
if (mConfig.show4gForLte) {
 mNetTIL.put(TM.NETWORK_TYPE_LTE, TI.FOUR_G);
} else {
 mNetTIL.put(TM.NETWORK_TYPE_LTE, TI.LTE);
}
mNetTIL.put(TM.NETWORK_TYPE_IWLAN, TI.WFC);
...
\end{lstlisting}
\end{lrbox}

\newsavebox\androidafter
\begin{lrbox}{\androidafter}
%\hspace{-50pt}
\begin{lstlisting}[escapechar=|,language=Java,basicstyle=\scriptsize\ttfamily,numbers=left,numbersep=5pt, numberstyle=\scriptsize\color{black},frame=None,xleftmargin=15pt,xrightmargin=2pt, aboveskip=5pt,belowskip=5pt]
...
 if (mConfig.show4gForLte) {
   mNetTIL.put(TM.NETWORK_TYPE_LTE, TI.FOUR_G);
+  if (mConfig.hideLtePlus) {|\label{lst:change1start}|
+   mNetTIL.put(TM.NETWORK_TYPE_LTE_CA, TI.FOUR_G);
+  } else {
+   mNetTIL.put(TM.NETWORK_TYPE_LTE_CA, TI.FOUR_G_PLUS);
+  }|\label{lst:change1end}|
  } else {
   mNetTIL.put(TM.NETWORK_TYPE_LTE, TI.LTE);
+  if (mConfig.hideLtePlus) {|\label{lst:change2start}|
+   mNetTIL.put(TM.NETWORK_TYPE_LTE_CA, TI.LTE);
+  } else {
+   mNetTIL.put(TM.NETWORK_TYPE_LTE_CA, TI.LTE_PLUS);
+  }|\label{lst:change2end}|
  }
  mNetTIL.put(TM.NETWORK_TYPE_IWLAN, TI.WFC);
...
\end{lstlisting}
\end{lrbox}

\newsavebox\lineageconflict
\begin{lrbox}{\lineageconflict}
\hspace{-5pt}
\begin{lstlisting}[escapechar=|,language=Java,basicstyle=\scriptsize\ttfamily,numbers=left,numbersep=5pt, numberstyle=\scriptsize\color{black},frame=None,xleftmargin=15pt,xrightmargin=2pt, aboveskip=5pt,belowskip=5pt]
...
if (mConfig.show4gForLte) {
<<<<<<< HEAD
 if (mContext.getResources()|\label{lst:lineageoscstart}|
  .getBoolean(R.bool.show_4glte_icon_for_lte)) {
  mNetTIL.put(TM.NETWORK_TYPE_LTE, TI.FOUR_G_LTE);
 } else if (mContext.getResources()
  .getBoolean(R.bool.show_network_indicators)) {
  mNetTIL.put(TM.NETWORK_TYPE_LTE, TI.LTE);
 } else {
  mNetTIL.put(TM.NETWORK_TYPE_LTE, TI.FOUR_G);
 }
 mNetTIL.put(TM.NETWORK_TYPE_LTE_CA, TI.FOUR_G_PLUS);
} else {
 mNetTIL.put(TM.NETWORK_TYPE_LTE, TI.LTE);
 if (mContext.getResources()
  .getBoolean(R.bool.show_network_indicators)){
  mNetTIL.put(TM.NETWORK_TYPE_LTE_CA, TI.FOUR_G_PLUS);
 } else {
  mNetTIL.put(TM.NETWORK_TYPE_LTE_CA, TI.LTE);|\label{lst:lineageoscend}|
=======
 mNetTIL.put(TM.NETWORK_TYPE_LTE, TI.FOUR_G);|\label{lst:androidcstart}|
 if (mConfig.hideLtePlus) {
  mNetTIL.put(TM.NETWORK_TYPE_LTE_CA, TI.FOUR_G);
 } else {
  mNetTIL.put(TM.NETWORK_TYPE_LTE_CA, TI.FOUR_G_PLUS);
 }
} else {
 mNetTIL.put(TM.NETWORK_TYPE_LTE, TI.LTE);
 if (mConfig.hideLtePlus) {
  mNetTIL.put(TM.NETWORK_TYPE_LTE_CA, TI.LTE);
 } else {
  mNetTIL.put(TM.NETWORK_TYPE_LTE_CA, TI.LTE_PLUS);|\label{lst:androidcend}|
>>>>>>> android/nougat-mr1.6-release
 }
}
...
\end{lstlisting}
\end{lrbox}

\newsavebox\lineageresolution
\begin{lrbox}{\lineageresolution}
%\hspace{-50pt}
\begin{lstlisting}[escapechar=|,language=Java,basicstyle=\scriptsize\ttfamily,numbers=left,numbersep=5pt, numberstyle=\scriptsize\color{black},frame=None,xleftmargin=15pt,xrightmargin=2pt, aboveskip=5pt,belowskip=5pt]
...
if (mConfig.show4gForLte) {
 if (mContext.getResources()
  .getBoolean(R.bool.show_4glte_icon_for_lte)) {
  mNetTIL.put(TM.NETWORK_TYPE_LTE, TI.FOUR_G_LTE);
 } else if (mContext.getResources()
  .getBoolean(R.bool.show_network_indicators)) {
  mNetTIL.put(TM.NETWORK_TYPE_LTE, TI.LTE);
 } else {
  mNetTIL.put(TM.NETWORK_TYPE_LTE, TI.FOUR_G);
  if (mConfig.hideLtePlus) {|\label{lst:los1start}|
   mNetTIL.put(TM.NETWORK_TYPE_LTE_CA, TI.FOUR_G);
  } else {
   mNetTIL.put(TM.NETWORK_TYPE_LTE_CA, TI.FOUR_G_PLUS);
  }|\label{lst:los1end}|
 }
 mNetTIL.put(TM.NETWORK_TYPE_LTE_CA, TI.FOUR_G_PLUS);
} else {
 mNetTIL.put(TM.NETWORK_TYPE_LTE, TI.LTE);
 if (mContext.getResources()
  .getBoolean(R.bool.show_network_indicators)){
  mNetTIL.put(TM.NETWORK_TYPE_LTE_CA, TI.FOUR_G_PLUS);
 } else {
  if (mConfig.hideLtePlus) {|\label{lst:los2start}|
   mNetTIL.put(TM.NETWORK_TYPE_LTE_CA, TI.LTE);
  } else {
   mNetTIL.put(TM.NETWORK_TYPE_LTE_CA, TI.LTE_PLUS);
  }|\label{lst:los2end}|
 }
}
...
\end{lstlisting}
\end{lrbox}

\begin{figure*}[!t]
\begin{minipage}[!t]{0.49\linewidth}
\vspace{-80pt}
\centering
\begingroup
\captionsetup[subfigure]{width=0.85\linewidth}
\vspace{10pt}
\subfloat[Part of the \texttt{mapIconSets} method in the Android OS before the update.\label{fig:androidbefore}]{
\usebox\androidbefore
}
\endgroup
\end{minipage}
\begin{minipage}[!t]{0.49\linewidth}
\centering
\begingroup
\captionsetup[subfigure]{width=0.85\linewidth}
\vspace{10pt}
\subfloat[Changes to \texttt{mapIconSets} in the Android OS after the update.\label{fig:androidafter}]{
\usebox\androidafter
}
\endgroup
\end{minipage}
\begin{minipage}[!t]{0.49\linewidth}
\vspace{-55pt}
\centering
\begingroup
\captionsetup[subfigure]{width=0.85\linewidth}
\vspace{10pt}
\subfloat[Merge conflict affecting \texttt{mapIconSets} in the \textsc{LineageOS}.\label{fig:lineageconflict}]{
\usebox\lineageconflict
}
\endgroup
\end{minipage}
\begin{minipage}[!t]{0.49\linewidth}
\vspace{-6pt}
\centering
\begingroup
\captionsetup[subfigure]{width=0.85\linewidth}
\vspace{10pt}
\subfloat[Conflict resolution in the \textsc{LineageOS}.\label{fig:lineageresolution}]{
\usebox\lineageresolution
}
\endgroup
\end{minipage}
\vspace{-10pt}
\caption{Example of a merge conflict and its resolution in the \textsc{LineageOS}.}
\vspace{-10pt}
\label{fig:motiv}
\end{figure*}

\section{Terminology \& Motivation}
\label{sec:motivation}

This section introduces some relevant terminology and presents an example that we use to motivate our work.

\subsection{Terminology}

Given a customized version of an OS, we call the project of the customized OS the \textit{downstream project} and the project of the original OS as the \textit{upstream project}. The downstream and the upstream projects are maintained in two different repositories, and we use the terms \textit{downstream repository} and \textit{upstream repository} to refer to these repositories. The downstream and the upstream projects use a version control system (VCS) to manage the changes associated with the files contained in the repositories. \textit{Downstream developers} make changes to the files in the downstream repository, while \textit{upstream developers} edit files in the upstream repository. Periodically, downstream developers pull changes from the upstream repository and merge them into the downstream repository. Downstream developers perform the \textit{merge operations} using the VCS. While performing merge operations, downstream developers can experience \textit{textual merge conflicts}. In this work, we use \textit{merge conflict} as an abbreviation for textual merge conflicts. A merge conflict can appear when the VCS cannot create a merged file given the changes to the file in the downstream and upstream repositories. Downstream developers resolve conflicts by suitably editing the conflicting changes associated with the file of the downstream repository.

\subsection{Motivation}

Fig.~\ref{fig:motiv} provides an example\footnote{We shorten variable names in the example due to space limitations.} of a merge conflict appearing in the \textsc{LineageOS}~\cite{lineageos}, an open-source customization of the Android OS that offers custom OS management and security features~\cite{thomas2015security, thomas2015lifetime}. In this example, the \textsc{LineageOS} is the downstream project and the Android OS is the upstream project. The conflict affects the method {\small\texttt{mapIconSets}} in the {\small\texttt{MobileSignalController.java}} file, which produces a mapping of data network types to icon groups. The conflict appeared when the downstream repository merged the changes from the upstream repository that have the {\small\texttt{nougat-mr1.6-release}} tag, which represents the changes characterizing a new version of the upstream repository. The conflict appears because the method {\small\texttt{mapIconSets}} in the downstream repository (lines~\ref{lst:lineageoscstart}-\ref{lst:lineageoscend} in Fig.~\ref{fig:lineageconflict}) has a different implementation with respect to the same method in the upstream repository (lines~\ref{lst:androidcstart}-\ref{lst:androidcend} in Fig.~\ref{fig:lineageconflict}). 

Downstream developers needed to handle such a conflict with careful attention because the part of the conflict from the upstream repository contains an update. Fig.~\ref{fig:androidbefore} and~\ref{fig:androidafter} provide the details of the update. Specifically, the method {\small\texttt{mapIconSets}} was updated by the upstream developers to handle new cases in the mapping of data network types to icon groups (lines~\ref{lst:change1start}-\ref{lst:change1end} and lines~\ref{lst:change2start}-\ref{lst:change2end} in Fig.~\ref{fig:androidafter}). Downstream developers identified these changes and suitably updated their implementation of the {\small\texttt{mapIconSets}} method as reported in Fig.~\ref{fig:lineageresolution} (lines~\ref{lst:los1start}-\ref{lst:los1end} and lines~\ref{lst:los2start}-\ref{lst:los2end}). 

This example presented how developers from \textsc{LineageOS} updated their downstream project to account for a change in the Android OS. In the rest of this paper, we present our study that characterizes whether and how developers from customizations of the Android OS update their downstream projects to account for the changes in the corresponding upstream projects.

\section{Study Design}
\label{sec:design}

%\subsection{Research Questions}

%Our study investigates how downstream customizations of the Android OS merge changes from their upstream OS.
In our study, we first aim to understand how often and how promptly developers from customized OSs merge changes from upstream repositories by answering the following research questions (RQs):

\begin{itemize}
    \setlength{\itemindent}{-10pt}
    \item[] \textbf{RQ1:} Do downstream developers merge changes from corresponding upstream projects?
    \item[] \textbf{RQ2:} What is the time lag between merge operations and corresponding commits in upstream repositories?
\end{itemize}

When merging changes from upstream projects, the merge operations might lead to conflicts that need to be explicitly resolved by the developers of the downstream projects.
We experimentally understand how often such conflicts happen and how they are resolved by answering the following RQs:

\begin{itemize}
    \setlength{\itemindent}{-10pt}
    \item[] \textbf{RQ3:} Do downstream developers experience conflicts when merging changes from upstream projects?
    \item[] \textbf{RQ4:} What types of code entities are affected by the merge conflicts?
    \item[] \textbf{RQ5:} What is the size of method-related conflicts?
    \item[] \textbf{RQ6:} How often do downstream developers change files in their projects when they experience a merge conflict?
    %\item \textbf{RQ7:} Do developers of Android apps use methods affected by merge conflicts?
\end{itemize}

Merge conflicts might indicate that customized projects might have diverged from their corresponding upstream project. Such differences may introduce compatibility issues into Android apps, which are developed based on a single framework version (often the official Android framework). We investigate the potential impact of merge conflicts with respect to compatibility issues by answering the following RQ:

\begin{itemize}
    \setlength{\itemindent}{-10pt}
    \item[] \textbf{RQ7:} Do developers of Android apps use methods affected by merge conflicts?
\end{itemize}

\subsection{Dataset Selection}

To identify relevant downstream projects, we analyzed a readily available and curated list of customized OSs based on Android~\cite{customizedandroid}. To the best of our knowledge, the list is an up-to-date list of custom OSs. The page is still under active maintenance and the latest update on the page was on March, 2022. At the time of writing, the list contains 33 projects. When we processed the list, we looked for projects that (i) customized the Android framework base, (ii) offered public access to the source code of their customization, and (iii) had their source code stored in repository using a version control system. Our study focuses on the Android framework base for two reasons. First, the Android framework base contains the implementation of core services in the OS. Second, we could readily investigate how the changes in this part of the OS affect Android apps, as apps heavily rely on this part of the framework. These selection criteria left us with nine candidate projects. Of the 24 projects we excluded from consideration, three projects were discontinued, and 21 projects did not provide access to the source code of the Android framework base. Among the remaining nine projects, we observed that one project (called \textsc{/e/}~\cite{eproject}) only performed one update to merge changes from the Android OS and decided to exclude the project from the study as it would have not significantly contributed to the results of our study. 

Table~\ref{tab:dataset} lists the remaining eight projects, which are the benchmarks we used in our study. The table reports the name of downstream project (column \textit{Downstream PN}), the name of the corresponding upstream project (column \textit{Upstream PN}), the link to the source code of the downstream project (column \textit{Repository URL}),
%and the year the projects were started (column \textit{Year Started}),
and the number of stars on Github (column \textit{\#. Stars}). All the downstream projects except \textsc{crdroidandroid} have the Android OS as their upstream. The upstream project for \textsc{crdroidandroid} is \textsc{LineageOS} and we decided to include this downstream project to also investigate the properties characterizing customized OSs deriving from customizations of the Android OS. To the best of our knowledge, these projects have been quite popular as they all have a significant number of stars on Github. (Replicant is not hosted on Github and, therefore, we could not collect the number of stars for the project.) Additionally, AOKP was used on more than 3.5 million devices in 2013~\cite{AOKP-users}.

\begin{table}[!t]
  	\begin{footnotesize}
  	\begin{center}
    \caption{Downstream projects considered in the study.}
    \vspace{-5pt}
    %\resizebox{\linewidth}{!}{
    \footnotesize
    \begin{tabular}{r |l|l|c}
        \hline
        \textit{Downstream PN} &  \textit{Upstream PN}  & \textit{Repository URL} & \textit{Stars} \\
        \hline
        \textsc{AOKP} & Android OS & \url{https://git.io/J3zId} & 31 \\
        \textsc{AOSPA} & Android OS & \url{https://git.io/JnXSE} & 96 \\
        \textsc{crdroidandroid} & LineageOS & \url{https://git.io/J3zLk} & 64 \\
        \textsc{LineageOS} & Android OS & \url{https://git.io/J3zLO} & 331 \\
        \textsc{OmniROM} & Android OS & \url{https://git.io/J3zLn} & 105 \\
        \textsc{replicant} & Android OS & \url{https://bit.ly/3gZkdoE} & - \\
        \textsc{ResurrectionRemix} & Android OS & \url{https://git.io/J3zLS} & 111 \\
        \textsc{SlimRoms}  &  Android OS & \url{https://git.io/J3zL5} & 40 \\
        \hline
    \end{tabular}
    \label{tab:dataset}
    \vspace{-18pt}
    \end{center}
    \end{footnotesize}
    %}
\end{table}

\section{Study Results}
\label{sec:result}

We answer the RQs of the study. For each RQ, we first describe the methodology used to answer the RQ and then present the results.

\subsection*{RQ1: \textit{Do downstream developers merge changes from corresponding upstream projects?}}

\noindent\textbf{Methodology:} To answer RQ1, we analyzed the version control history of the downstream projects and identified commits that merged updates from the corresponding upstream repositories. To automatically identify these commits, we first manually inspected the version control history of the repositories and identified that these commits have two parent commits (which was expected as the commits are merge commits) and the commit \textit{identifier} (also known as the commit \textit{hash}) of one of the two parents was also present in the corresponding upstream project. Moreover, some of the merge commits also contain commit messages describing the merge operation from the upstream, which we used to validate our manual inspection. Listing~\ref{lst:commitexample} provides an example of a relevant merge commit from the \textsc{LineageOS}.

\begin{lstlisting}[caption={Commit merging changes from the Android OS.},captionpos=b,abovecaptionskip=2pt,belowcaptionskip=0pt,label={lst:commitexample},escapechar=|,language=Java,basicstyle=\scriptsize\ttfamily,numbers=left,numbersep=5pt, numberstyle=\scriptsize\color{black},frame=None,xleftmargin=15pt,xrightmargin=2pt, aboveskip=8pt,belowskip=8pt]
commit 80d8b6467ec454e76eb69eb49f80f74d1e|\label{lst:mcid}|
Merge: 9b48a29cca52 37a24f52e6be|\label{lst:mcidparents}|
Author: xxx<xxx@xxx.com>
Date:   Sat Nov 16 08:46:50 2019 -0700
Merge android-10.0.0_r11 into lineage-17.0
Android 10.0.0 release 11
Change-Id: I05fb998d2e35db534880c89921f595d2225dc9a2
\end{lstlisting}

The listing reports the identifier for the merge commit ({\small\texttt{80d8b6467 ec4}} at line~\ref{lst:mcid}) and the identifiers associated with the parent commits (line~\ref{lst:mcidparents}). The first commit identifier ({\small\texttt{9b48a29cca52}}) across the two parents denotes a commit in the \textsc{LineageOS}, while the second commit identifier ({\small\texttt{37a24f52e6be}}) corresponds to the commit containing the changes that the downstream repository merged from the upstream repository. Given these characteristics, in our automated analysis, we identified relevant merge commits by checking whether a merge commit had one of the parent commits whose identifier also appeared in the corresponding upstream repository. After identifying a relevant merge commit, we also checked whether the parent commit from the upstream repository had a tag associated with the commit. We identified this information by retrieving the list of tags and corresponding commit identifiers from the upstream repository. We retrieve this information as tags identifying version releases in the Android OS. For example, the tag {\small\texttt{android-10.0.0\_r1}} in line 5 of Listing~\ref{lst:commitexample} represents a version of the Android system. For \textsc{crdroidandroid} we did not retrieve this information as its upstream repository (\textsc{LineageOS}) stopped using tags in 2015~\cite{lineageos-github}.

\begin{table}[!t]
    \begin{scriptsize}
  	\begin{center}
  	\footnotesize
    \caption{Merge operations in downstream projects.}
    \vspace{-5pt}
    \resizebox{\linewidth}{!}{
    \begin{tabular}{r|c|c|c|c|c }
    \hline
        \textit{\makecell{Downstream\\ Project\\ Name}}       &   \textit{\makecell{Total\\ Downstream\\ Commits}}  &  \textit{\makecell{Downstream\\ Specific\\ Commits}}  &   \textit{\makecell{Merge\\ Commits}}   &   \textit{Tags}   &  \textit{Tags (\%)}  \\
    \hline
        \textsc{AOKP}               & 374,786 &  4,716   &  15     &   14     &   2.50\%    \\
        \textsc{AOSPA}              & 512,846 &  12,370  &  256    &   53     &   9.46\%    \\
        \textsc{crdroidandroid}     & 517,912 &  8,517   &  351    &   -      &    -        \\
        \textsc{LineageOS}          & 521,435 &  20,539  &  163    &   136    &   24.29\%   \\
        \textsc{OmniROM}            & 498,703 &  5,707   &  251    &   92     &   16.43\%   \\
        \textsc{replicant}          & 213,805 &  6,937   &  34     &   22     &   3.93\%    \\
        \textsc{ResurrectionRemix}  & 441,335 &  14,585  &  45     &   37     &   6.61\%    \\
        \textsc{SlimRoms}           & 429,688 &  8,081   &  33     &   28     &   5.00\%    \\
    \hline
        \textit{Total}              & 3,510,510 &  81,452  & 1,148  &   382    &   - \\ % 68.22
    \hline
    \end{tabular}}
    \label{tab:google_android_tags}
    \vspace{-10pt}
    \end{center}
    \end{scriptsize}
\end{table}

\noindent\textbf{Results:} Table~\ref{tab:google_android_tags} reports the results of running our analysis on the eight downstream projects. For each of the downstream projects, the table provides the total number of commits in the downstream project (column \textit{Total Downstream Commits}), the number of commits specific to the downstream project (column \textit{Downstream Specific Commits}), the number of commits that merge updates from the upstream projects (column \textit{Merge Commits}), the number of commits having a tag (column \textit{Tags}), and the percentage of tags as compared to the total number of tags\footnote{We identified the tags from the Android OS documentation~\cite{androidtags}.} identifying a version release in the corresponding upstream project (column \textit{Tags (\%)}). The difference between the total number of downstream commits and the number of downstream-specific commits is due to the fact that merge operations bring a large number of commits from upstream repositories into downstream repositories. The number of downstream-specific commits also shows that the projects are characterized by substantial development.

The results reported in Table~\ref{tab:google_android_tags} show that \textbf{there is a high variance in the number of merge operations across the downstream projects}. For example, developers from \textsc{LineageOS} performed 163 merge operations, while developers from \textsc{AOKP} performed this type of operations only 15 times, even if the projects started similar times (2009 for \textsc{LineageOS} and 2011 for \textsc{AOKP}) and are still active. The percentage of tags associated with the merge operations also varies significantly. For example, in \textsc{AOKP}, 93.3\% (14/15) of the merge commits have a tag associated with them, while in \textsc{AOSPA}, only 20.7\% (53/256) of the commits are associated with a tag. Overall, the downstream projects performed 1,148 merge commits to merge updates from their upstream repositories, but these commits do not always have a tag associated with them. In fact, only 47.9\% of the commits are associated with a tag, meaning that in roughly half of the cases, the commits are not associated with a new version release of the upstream project but some intermediate version of the upstream project. We believe that this result paves the way for new research on automatically identifying and suggesting to developers which updates to merge in their downstream projects. We believe that automated techniques based on static analysis and machine learning could compare downstream and upstream changes to identify the proper time to merge updates from upstream repositories.

\begin{figure}[!t]
    \centering
    \begin{minipage}[b]{0.98\linewidth}
         \centering
         \includegraphics[trim={0 0 0 0.85cm},clip,width=\textwidth]{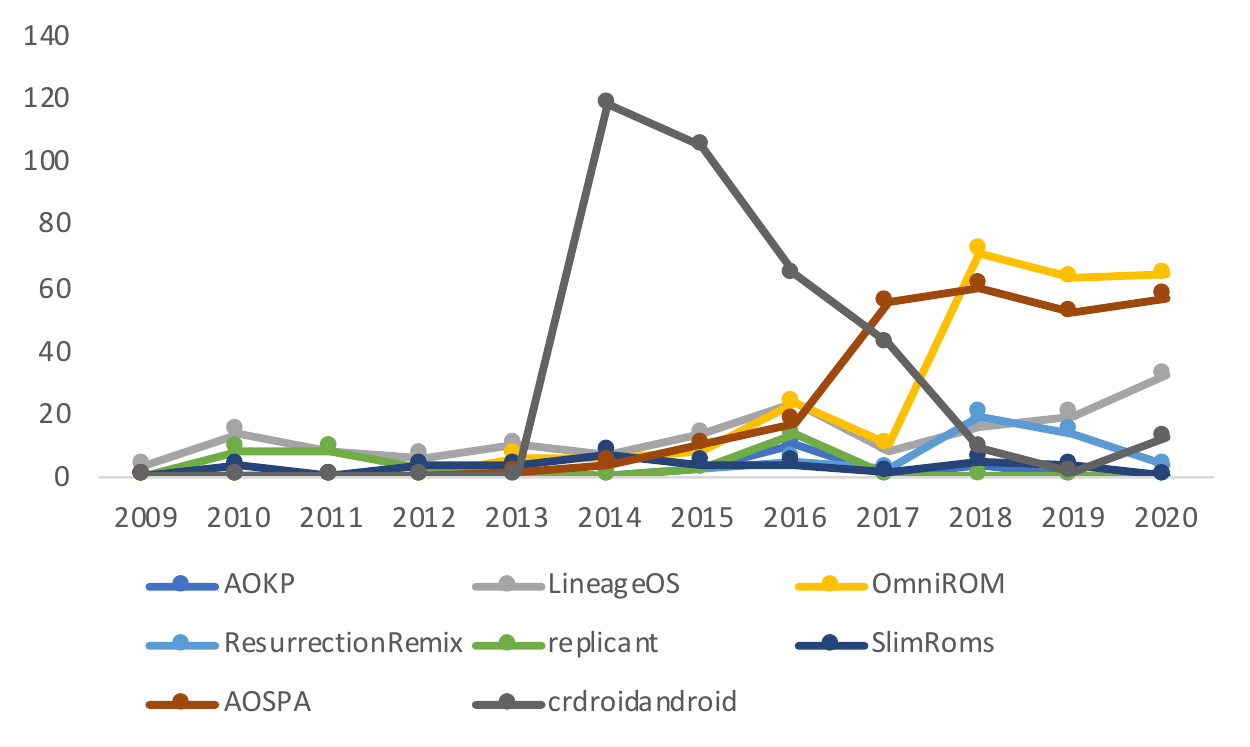}
         \vspace{-22pt}
         \caption{Distribution of merge commits over time.}
         \label{fig:merge_dist}
         \vspace{-15pt}
    \end{minipage}
\end{figure}

The results in Table~\ref{tab:google_android_tags} also show that \textbf{downstream projects merge updates only for a small percentage (9.7\%) of the versions released by the upstream projects}. 
%mf: TSE reviewers mentioned that this claim is very vague and O decided to remove it 
%This result shows that this maintenance task is likely to have a high cost for developers and that downstream developers need additional automated or semi-automated techniques to perform the task.
Fig.~\ref{fig:merge_dist} provides a plot of the merge operations over time. The plot reports the number of merge commits performed by each of the downstream projects in the years from 2009 (the year the first downstream project was created) to 2020 (the year we started this study). All projects have merged updates from their upstream projects for at least three consecutive years. Furthermore, five projects have never stopped merging updates since they were created. The plot also shows that \textsc{AOSPA}, \textsc{OmniROM}, and \textsc{LineageOS} are the projects with the highest number of merge operations in the recent years.

\begin{figure}[t!]
    \centering
    \begin{minipage}[b]{0.98\linewidth}
         \centering
         \includegraphics[trim={0 0 0 0.85cm},clip,width=\textwidth]{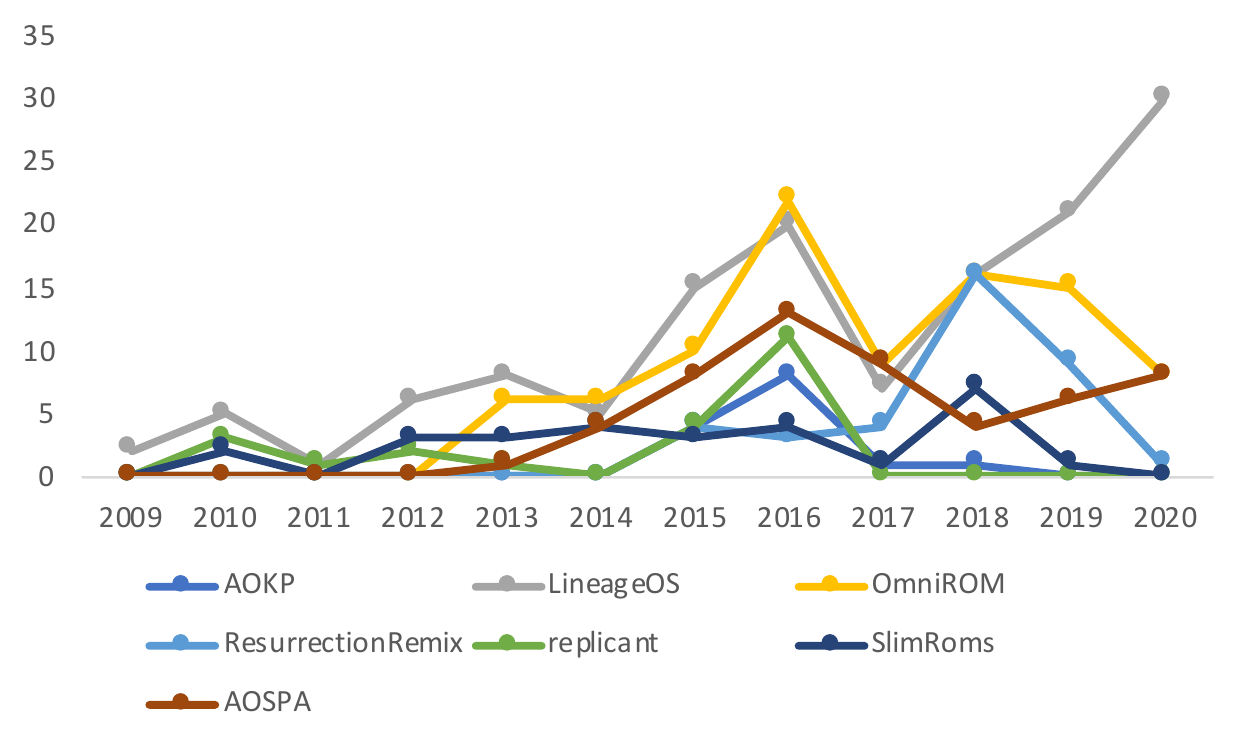}
         \vspace{-22pt}
         \caption{Merge commits with tags over time.}
         \label{fig:tag_dist}
         \vspace{-15pt}
    \end{minipage}
\end{figure}

To further detail how developers have merged updates over time, Fig.~\ref{fig:tag_dist} plots the number of merge commits that have an associated tag over time. The plot reveals that \textbf{downstream projects do not consistently merge updates from new releases of the upstream projects}. However, the \textsc{LineageOS} is increasing the number of this type of merge commits in recent years. Additionally, after a manual inspection of the tags, we did not identify any patterns in the tags that are merged by any of the projects. For instance, the projects did not always merge the first or the last release tag associated with a major version release of the corresponding upstream project. Android has 18 major releases to date~\cite{androidtags}. We expected to see these tags used frequently as they indicate a new major revision of the official upstream repository or the latest update, which could provide long term support. Overall, the plots seem to indicate that, currently, downstream projects do not use a systematic approach (periodically merge or merge as soon as a new major version is released) to determine when to merge updates from upstream repositories.

\begin{tcolorbox}[before skip=0.4cm, after skip=0.6cm, title=\textbf{RQ1 Findings}, left=2pt, right=2pt,top=2pt,bottom=2pt]
Downstream projects merge updates from upstream projects. The merge operations are both based on commits identifying version releases in the upstream projects (47.9\% of the case) but also on other commits (52.1\% of the cases). Additionally, downstream projects perform merge operations only for a small portion (9.9\%) of all the version releases in the upstream projects. Finally, the results seem to indicate that downstream projects do not use a systematic approach to decide which updates to merge from upstream repositories.
\end{tcolorbox}

\subsection*{RQ2: \textit{What is the time lag between merge operations and corresponding commits in upstream repositories?}}

\noindent\textbf{Methodology:} For every relevant merge commit identified in RQ1, we computed the time lag between the time the merge commit was performed and the time when the associated parent commit was made in the upstream repository. Considering the example in Listing~\ref{lst:commitexample}, the analysis computed the time lag between commit {\small\texttt{80d8b6467ec4}} and commit {\small\texttt{37a24f52e6be}}.

\begin{figure}[h!]
    \centering
    \includegraphics[width=0.75\linewidth]{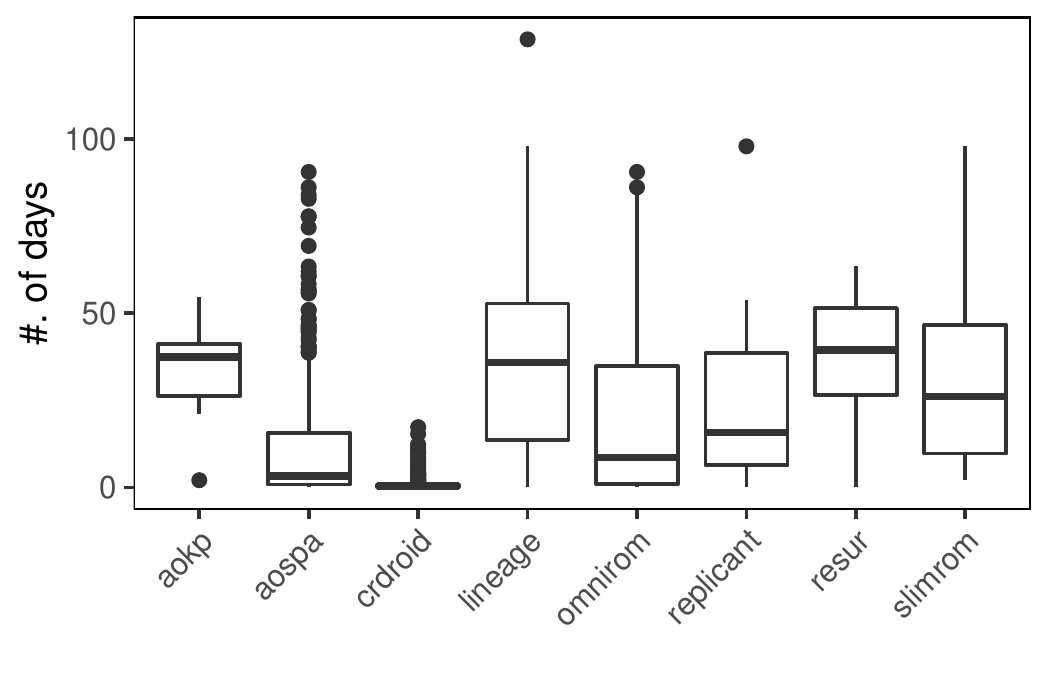}
    \vspace{-24pt}
    \caption{The distribution of merge time lag.}
    \label{fig:merge_time_lag}
    \vspace{-15pt}
\end{figure}

\noindent\textbf{Results:} Table~\ref{tab:merge_commit_avg} reports our analysis results. Specifically, for each of the downstream projects, the table reports the number of merge commits considered (\textit{Merge Commits}), the average time lag in days between the upstream commit associated with a merge operation and the related downstream merge commit (\textit{Time Lag}), and the average time difference in days between downstream-specific commits (\textit{Downstream Commit}). Fig.~\ref{fig:merge_time_lag} provides more details on the time lag and shows its distribution for all the projects. \textsc{crdroidandroid} has an average time lag of 0.96 days and median time lag of 0.40 days and it is the shortest average time lag and median time lag across all projects. This very short time lag seem to indicate that \textbf{maintainers from this project closely follow the changes in the upstream repository}, and when the maintainers see relevant changes, they try to merge the changes immediately. \textsc{ResurrectionRemix}, instead, is characterized by the longest average time lag and highest median time lag, which is equal to 37.60 days and 39.51 days, respectively. Excluding \textsc{crdroidandroid}, all projects have an average lag time of more then 10 days and for most of the projects (five) the lag time is above 20 days. Moreover, some of the merge operations are even finished more than three months later, as shown in Fig.~\ref{fig:merge_time_lag} for projects \textsc{AOSPA} and \textsc{LineageOS}. The average time lag is in stark contrast with the time difference between downstream-specific commits, which for all projects is less than a day. This result reveals that \textbf{the time lag is a critical aspect characterizing downstream projects}. In fact, the long time lag might negatively impact downstream projects as they might see critical bug fixes and security updates only after several days or weeks. This situation might expose user devices running the customized OSs to security attacks.

\begin{tcolorbox}[before skip=0.4cm, after skip=0.6cm, title=\textbf{RQ2 Findings}, left=2pt, right=2pt,top=2pt,bottom=2pt]
Most of the downstream projects take more than 20 days to bring changes from their corresponding upstream projects. Because security patches might not reach downstream projects in a timely manner, user devices running customized OSs might often be exposed to security vulnerabilities.
\end{tcolorbox}

\begin{table}[!t]
    \begin{scriptsize}
  	\begin{center}
    \caption{Average lag time associated with merge commits.}
    \vspace{-10pt}
    %\resizebox{\linewidth}{!}{
    \begin{tabular}[\linewidth]{r|c|c|c}
    \hline
        \textit{Downstream PN}            &  \textit{Merge Commits}  & \textit{\makecell{Time Lag\\(days)}}  & \textit{\makecell{Downstream Commit\\(days)}}  \\ % \hline & \multicolumn{2}{c}{\#. commits (in subsequent merges)} \\ \hline
            % &   &   Lag \\ % & Resolve  & Downstream & Upstream \\
    \hline
        \textsc{AOKP}               &    15    &   33.70   & 0.42  \\ % & 0.83  &   649.33      &    428.07   \\
        \textsc{AOSPA}              &    256   &   13.10   & 0.42  \\ % & 3.60  &   722.19      &    749.125  \\
        \textsc{crdroidandroid}     &    351   &   0.96    & 0.28  \\ % & 3.23  &   30.04       &    26.38    \\
        \textsc{LineageOS}          &    163   &   33.58   & 0.33  \\ % & 5.69  &   654.31      &    372.55   \\
        \textsc{OmniROM}            &    251   &   18.25   & 0.52  \\ % & 3.43  &   773.34      &    4,250,46 \\
        \textsc{replicant}          &    34    &   22.55   & 0.78  \\ % & 0.74  &   771.26      &    1,791.47 \\
        \textsc{ResurrectionRemix}  &    45    &   37.60   & 0.24  \\ % & 3.17  &   484.73      &    124.16   \\
        \textsc{SlimRoms}           &    33    &   31.20   & 0.67  \\ % & 15.03  &   1,595.15    &    2,650.18 \\
    \hline
        \textit{Total/Average}      &   1,148  &   23.87   & 0.46 \\
    \hline
    \end{tabular}%}
    \label{tab:merge_commit_avg}
    \vspace{-20pt}
    \end{center}
    \end{scriptsize}
\end{table}

\subsection*{RQ3: \textit{Do downstream developers experience conflicts when merging changes from upstream projects?}}

\begin{table}[t!]
    \begin{footnotesize}
  	\begin{center}
    \caption{Conflicts when merging changes from upstream.}
    \vspace{-4pt}
    \begin{tabular}[0.96\linewidth]{r|c|c}
    \hline
        \textit{Downstream PN}    &  \textit{Merge Commits} & \textit{Conflicts} \\
        \hline
        \textsc{AOKP}               & 15           & 6   (40.0\%)   \\
        \textsc{AOSPA}              & 256          & 138 (53.91\%)  \\
        \textsc{crdroidandroid}     & 351          & 78  (22.22\%)  \\
        \textsc{LineageOS}          & 163          & 70  (42.94\%)  \\
        \textsc{OmniROM}            & 251          & 128 (51.0\%)   \\
        \textsc{replicant}          & 34           & 18  (52.94\%)  \\
        \textsc{ResurrectionRemix}  & 45           & 16  (35.56\%)  \\
        \textsc{SlimRoms}           & 33           & 20  (60.61\%)  \\
        % /e/                & 189    & 1   (0.53\%)   \\
        \hline
        \textit{Total}             &  1,148        & 474 (41.29\%) \\
        \hline
    \end{tabular}
    \label{tab:total_conf_merge}
    \vspace{-15pt}
    \end{center}
    \end{footnotesize}
\end{table}

\noindent\textbf{Methodology:}
%This and the following RQs focus on merge conflicts as they have shown to be one of the primary challenges in the development of divergent software systems~\cite{sung2020towards}.
To identify conflicts, we processed the merge commits from RQ1, and for each commit, we use the information contained in the commit to (i) reset the state of the downstream repository to the parent commit from the downstream repository, (ii) reset the state of the upstream repository to the parent commit from the upstream repository, and (iii) perform the merge operation locally. In these steps, we reset the state of a repository using the {\small\texttt{git reset --hard <commit-id>}} command and we perform the merge operation by first adding the upstream repository as a remote repository of the downstream repository (using the {\small\texttt{git remote add <remote-name> <upstream-repository-directory>}} command), and then actually performing the merge operation (using the {\small\texttt{git merge <remote-name>/<branch>}} command). After we performed these operations for each of the merge commits, we counted the number of commits that lead to at least one merge conflict. To give an example, when we processed the merge commit reported in Listing~\ref{lst:commitexample}, we reset the state of the downstream repository to commit {\small\texttt{9b48a29cca52}}, reset the state of the upstream repository to commit {\small\texttt{37a24f52e6be}}, and identified a conflict after we performed the merge operation.
%mf: removed for space
%The conflict is reported in Listing~\ref{lst:local_merge_example} and affects the {\small\texttt{PowerManagerService.java}} file\todo{This example could god if we need space}.
% \begin{lstlisting}[caption={Example of conflict.},captionpos=b,abovecaptionskip=2pt,belowcaptionskip=0pt,label={lst:local_merge_example},escapechar=|,language=Java,basicstyle=\scriptsize\ttfamily,numbers=left,numbersep=5pt, numberstyle=\scriptsize\color{black},frame=None,xleftmargin=15pt,xrightmargin=2pt, aboveskip=8pt,belowskip=8pt]
% Auto-merging .../server/power/PowerManagerService.java
% CONFLICT (content): Merge conflict in
% .../server/power/PowerManagerService.java
% \end{lstlisting}
% \vspace{-5pt}

\noindent\textbf{Results:} Table~\ref{tab:total_conf_merge} reports the number of conflicts that we identified in our analysis. Specifically, for each of the downstream projects, the table reports the number of merge commits considered (column \textit{Merge Commits}) and the number of commits affected by a conflict (column \textit{Conflicts}). Across all projects, \textbf{41.3\% of the total number of 1,148 commits are affected by a conflict} and all the projects are characterized by at least one conflict. The project with the highest percentage of commits affected by conflict is \textsc{SlimRoms} while the project with the lowest percentage is \textsc{crdroidandroid}. Even if \textsc{crdroidandroid} has the lowest percentage, the project still encounters a considerable number of conflicts. These results show that \textbf{conflicts are an important aspect in the development of customized OSs}.
%mf: removed as it is vague and TSE reviewers did not like it
%and further research on automated or semi-automated techniques to resolve the conflicts is needed.

\vspace{-5pt}
\begin{tcolorbox}[before skip=0.4cm, after skip=0.6cm, title=\textbf{RQ3 Findings}, left=2pt, right=2pt,top=2pt,bottom=2pt]
Even though downstream projects do not experience a conflict in every merge operation, all of the projects experience conflicts, and most of the projects (six) experience a conflict 40\% of the time.
%mf: removed as it is vague and TSE reviewers did not like it
%The overall high number of conflicts and the fact that every project experiences a considerable number of conflicts motivates future research on techniques to help developers resolve conflicts automatically or semi-automatically.
\end{tcolorbox}

\begin{table}[t!]
    \centering
    \scriptsize
    \caption{Files affected by merge conflicts.}
    \vspace{-4pt}
    \resizebox{\linewidth}{!}{
    \begin{tabular}{r|c|c|c|c|c|c|c}
    \hline
    \multirow{2}{8em}{\textit{Downstream Project Name}}   &  \multicolumn{6}{c|}{\textit{Files}}  &  \multirow{2}{2em}{\textit{Project Tot.}} \\
    \cline{2-7}
                               &  Java  &  aidl  &  cpp  &  xml  &  png  &  other  &       \\
    \hline
    \textsc{AOKP}              &  49    &  2     &   1   &   8   &   0   &   0   &  60    \\
    \textsc{AOSPA}             &  943   &  0     &  47   &  498  &  45   &   40  &  1,573 \\
    \textsc{crdroidandroid}    &  116   &  5     &  1    &  23   &   2   &   2   &  149   \\
    \textsc{LineageOS}         &  526   &  25    &  37   &  173  & 104   &   24  &  889   \\
    \textsc{OmniROM}           &  934   &  32    &  35   &  454  &  53   &   42  &  1,550 \\
    \textsc{replicant}         &  119   &  3     &  12   &  44   &  14   &   13  &  205   \\
    \textsc{ResurrectionRemix} &  204   &  8     &  8    &  46   &   0   &   4   &  272   \\
    \textsc{SlimRoms}          &  179   &  15    &  11   &  55   &  139  &   2   &  401   \\
    \hline
    \textit{File Type Total}  &  3,070   &  90    &  152   &  1,301   &  357  &   127   &  5,099   \\
    \hline
    \end{tabular}}
    \label{tab:merge_files}
    \vspace{-10pt}
\end{table}

\subsection*{RQ4: \textit{What types of code entities are affected by the merge conflicts?}}

\begin{table*}[t!]
    \centering
    \scriptsize
    \caption{Characterization of files and code entities affected by merge conflicts.}
    \vspace{-10pt}
    \resizebox{\linewidth}{!}{
    \begin{tabular}[0.98\linewidth]{r|c|c|c|c|c|c|c|c|c|c|c|c|c|c}
    \hline
        & \multicolumn{2}{c|}{\textit{Merge Ops}} & \multicolumn{1}{c|}{\textit{Files}} & \multicolumn{11}{c}{\textit{Code Entities}}
    \\ \hline
        \textit{Downstream PN} & \textit{Total} & \textit{Conflicts} & \textit{Java} & \textit{Import} & \textit{Class} & \textit{Field} &  \textit{Constructor} & \textit{Method} & \textit{Enum} & \textit{\makecell[c]{Class\\ Comment}} & \textit{\makecell[c]{Field\\ Comment}} & \textit{\makecell[c]{Constructor\\ Comment}} & \textit{\makecell[c]{Method\\ Comment}} & \textit{\makecell[c]{Comment}} \\
    \hline
        \textsc{AOKP}              & 15  & 6 (40.0\%)    & 49 (81.67\%)  &  4   & 1 & 20  & 3   & 36  & 0 & 0 & 8  & 0 & 1  & 5   \\
        \textsc{AOSPA}             & 256 & 138 (59.95\%) & 943 (62.32\%) &  124 & 5 & 218 & 102 & 836 & 0 & 3 & 93 & 3 & 62 & 241 \\
        \textsc{crdroidandroid}    & 351 & 78 (22.33\%)  & 116 (77.85\%) &  30  & 1 & 30  & 9   & 90  & 0 & 1 & 17 & 0 & 1  & 14  \\
        \textsc{LineageOS}         & 163 & 70 (42.94\%)  & 526 (59.17\%) &  106 & 6 & 175 & 55  & 576 & 1 & 6 & 48 & 0 & 33 & 83  \\
        \textsc{OmniROM}           & 251 & 128 (51.0\%)  & 934 (60.26\%) &  156 & 8 & 223 & 104 & 876 & 0 & 5 & 93 & 3 & 64 & 178 \\
        \textsc{replicant}         & 34  & 18 (52.94\%)  & 119 (58.05\%) &  15  & 2 & 31  & 11  & 112 & 0 & 1 & 15 & 0 & 6  & 11  \\
        \textsc{ResurrectionRemix} & 45  & 16 (35.56\%)  & 204 (75.00\%) &  55  & 1 & 75  & 36  & 217 & 0 & 0 & 17 & 0 & 8  & 38  \\
        \textsc{SlimRoms}          & 33  & 20 (60.61\%)  & 179 (44.64\%) &  34  & 1 & 71  & 16  & 198 & 1 & 0 & 18 & 0 & 15 & 39  \\
    \hline
        \textit{Total}          & 1,148  & 474 (41.29\%) & 3,070 (60.92\%) &  524 & 25 & 843 & 336 & 2,941 & 2 & 16 & 309 & 6 & 190 & 609 \\
        % /e/ & 1 & 1 & 0 & 0 & 0 & 0 & 0 & 0 & 0 & 0 & 0 & 0 & 0 \\
    \hline
    \end{tabular}}
    \label{tab:merge_conf_types}
    \vspace{-5pt}
\end{table*}

\begin{figure}[h!]
    \centering
    % \includegraphics{}
    % \caption{Caption}
    % \label{fig:my_label}
    \vspace{-5pt}
    \hspace*{-30pt}   
    \begin{minipage}[t]{.45\textwidth}
    \scriptsize
    \begin{lstlisting}[caption={Code snippet before update.},captionpos=b,abovecaptionskip=2pt,belowcaptionskip=0pt,label={lst:cmt_merge_before},escapechar=|,language=Java,basicstyle=\scriptsize\ttfamily,numbers=left,numbersep=5pt, numberstyle=\scriptsize\color{black},frame=None,xleftmargin=15pt,xrightmargin=2pt, aboveskip=8pt,belowskip=8pt]
/**
  * <p>This can be used when the request is unnecessary 
  * or will be superceeded by a request that
  * will soon be queued. 
  *
  * @return the future id of the canceled request, 
  * or {@link FillRequest#INVALID_REQUEST_ID} if
  * no {@link PendingFillRequest} was canceled.
  */
public CompletableFuture<Integer> cancelCurrentRequest(){
  return CompletableFuture.supplyAsync(() -> {
    if (isDestroyed()) { return INVALID_REQUEST_ID;}
    BasePendingRequest<RemoteFillService, 
      IAutoFillService> canceledRequest = 
      handleCancelPendingRequest();
      return canceledRequest instanceof 
       PendingFillRequest ? ((PendingFillRequest) 
         canceledRequest).mRequest.getId() 
         : INVALID_REQUEST_ID; }, mHandler::post);
}
    \end{lstlisting}
    \end{minipage}
    \centering
    \vspace{-10pt}
    \hspace*{-30pt}
    \begin{minipage}[t]{.45\textwidth}
    \scriptsize
    \vspace{-15pt}
    \begin{lstlisting}[caption={Code snippet after update that led to a conflict.},captionpos=b,abovecaptionskip=2pt,belowcaptionskip=0pt,label={lst:cmt_merge_after},escapechar=|,language=Java,basicstyle=\scriptsize\ttfamily,numbers=left,numbersep=5pt, numberstyle=\scriptsize\color{black},frame=None,xleftmargin=15pt,xrightmargin=2pt, aboveskip=8pt,belowskip=8pt,firstnumber=1]{Name}
/**
  * <p>This can be used when the request is unnecessary 
  * or will be superceeded by a request that
  * will soon be queued.
  *
  * @return the id of the canceled request, 
  * or {@link FillRequest#INVALID_REQUEST_ID} if
  * no {@link FillRequest} was canceled.
  */
public int cancelCurrentRequest() {
  synchronized (mLock) {
    return mPendingFillRequest != null && 
    mPendingFillRequest.cancel(false) ?  
      mPendingFillRequestId : INVALID_REQUEST_ID;
  }}
    \end{lstlisting}
    \end{minipage}
\end{figure}

\noindent\textbf{Methodology:} To identify code entities affected by the merge conflicts, we processed all the conflicts we identified in RQ3, and performed a two steps analysis. First, we identified the files affected by conflicts and categorized the files based on their types. Second, for all the Java files (which are the predominant type of source code file in the part of the OS we analyzed), we also identified the types of code entities (\eg import statements, class fields, method bodies, \etc) affected by the conflicts. In the analysis, we considered code entities so that we could characterize Java files in their entirety. To perform this second step, we map the conflict information to the abstract syntax tree (AST) of the Java files as they appear in the downstream repository before performing the merge operation.

\noindent\textbf{Results:} Tables~\ref{tab:merge_files} and~\ref{tab:merge_conf_types} report the results of our analysis. Table~\ref{tab:merge_files} reports summary information describing the file types affected by the conflicts. Specifically, for each downstream project, the table reports the number of files of a certain file type that have at least one conflict. \textbf{Most of the time the conflicts involve Java files}. Specifically, for all projects, except for \textsc{SlimRoms}, more than 55\% of the files that have a conflict are Java files. The remaining portion of files contains aidl, cpp, xml, and png files, and also additional file types with included configuration and documentation files.

Table~\ref{tab:merge_conf_types} provides details on the code entities affected by conflicts. In the table, column \textit{Merge Ops} reports the total number of merge operations analyzed and the portion of operations leading to a conflict, column \textit{Files} presents the number of Java files affected by at least one conflict, and column \textit{Code Entities} details the number of code entities affected by a conflict divided by the entity type. The \textbf{entities most affected by conflicts are methods, followed by fields}. This result was expected as methods and fields are easy to leverage for extending downstream projects with new functionalities. It is worth noting that a considerable number of conflicts are caused by comments and import statements.
This result suggests that the set of classes defined in the downstream projects is likely to be different as compared to upstream projects, which indicates that resolving the changes from the upstream projects might not be always an easy task. The last five columns in Table~\ref{tab:merge_conf_types} summarize the number of conflicts caused by the comments associated with classes, fields, constructors, methods, and inline comments or stand-alone comments not specific to any code.
This part of Table~\ref{tab:merge_conf_types} shows \textbf{that not only executable code but also comments can cause conflicts} when merging changes from upstream to downstream repositories. After manually inspecting the conflicts associated with method comments, we identified that some of the conflicts are due to changed method signatures and some are associated with semantic differences between the code in downstream and upstream repositories.

Listings~\ref{lst:cmt_merge_before} and~\ref{lst:cmt_merge_after} represent an example of a method comment that lead to a conflict. The method {\small\texttt{cancelCurrentRequest}} returns an integer value representing the identifier of a canceled request. Before resulting in a conflict, the method returned the future identifier through an integer wrapper class of {\small\texttt{CompletableFuture}}. After changing the code in the downstream repository, the method implementation was updated to return a primitive integer value. It is worth noting that the updated implementation leads to the change of the comment associated with the method. This change highlights that it might be possible to have code changes that are complex and involving semantic changes. Although the number of comments causing conflicts is smaller than the number of conflicts affecting executable code, \textbf{the difference in the comments suggests that some code changes are critical and hence have to be properly resolved by downstream projects}, like the ones highlighted in the motivating example of Section~\ref{sec:motivation}. If downstream developers do not properly handle updates to existing methods, user apps running on customized OSs might experience compatibility issues.

% specification:
% Merge: the number of conflict merges
% Conf.: the number of conflict files
% Java: the number of conflict java files
% Import: conflict happens at import statement
% Class Dec: conflict happens at Class Declaration
% Filed: conflict happens at field
% Constructor: conflict happens at constructor
% Method: conflict happens at Methods
% Enum: conflict happens at Enum declaration
% ClassC: conflict happens at comments for whole class
% FiledC: conflict happens at comment for field
% Const.C: conflict happens at comment for constructor
% MethodC: conflict happens at comment for method
% EnumC: conflict happens at comment for Enum.

\vspace{-6pt}
\begin{tcolorbox}[before skip=0.4cm, after skip=0.6cm, title=\textbf{RQ4 Findings}, left=2pt, right=2pt,top=2pt,bottom=2pt]
The projects we considered experienced the majority of the conflicts in Java files. The most affected types of code entities are methods and fields. Our analysis also identified that code comments have conflicts, highlighting the possibility that the corresponding code conflicts might involve semantic changes.
%mf: removed as it is very vague and TSE reviewers did not like it.
%Finally, automated techniques aiming to help downstream developers in resolving code conflicts should pay special attention to conflicts characterizing methods and fields.
\end{tcolorbox}

\vspace{-2pt}
\subsection*{RQ5: \textit{What is the size of method-related conflicts?}}

%In the analysis of RQ4, we identified that most of the conflicts affect methods, with a total of 2,941 method-related conflicts across the eight downstream projects.
\noindent\textbf{Methodology:} In this research question, we compute the size of the conflicts affecting code methods to estimate the cost of analyzing and resolving the conflicts. We compute the size of the conflicts by looking at the number of lines and AST nodes involved in the conflicts. To compute the number of AST nodes involved in the conflicts, we leverage an AST parser~\cite{javaparser} to analyze the portions of the Java source code files affected by the conflicts. Specifically, we first scan all the Java source code files affected by a conflict and then map AST nodes to the lines of the conflicts in the relevant files of the downstream and upstream projects. After that, we determine the size of the conflict by computing the number of nodes actually involved in lines affected by the conflicts.

\begin{figure}[t!]
    \centering
    \includegraphics[trim={0 0 0.4cm 0},clip,width=0.95\linewidth]{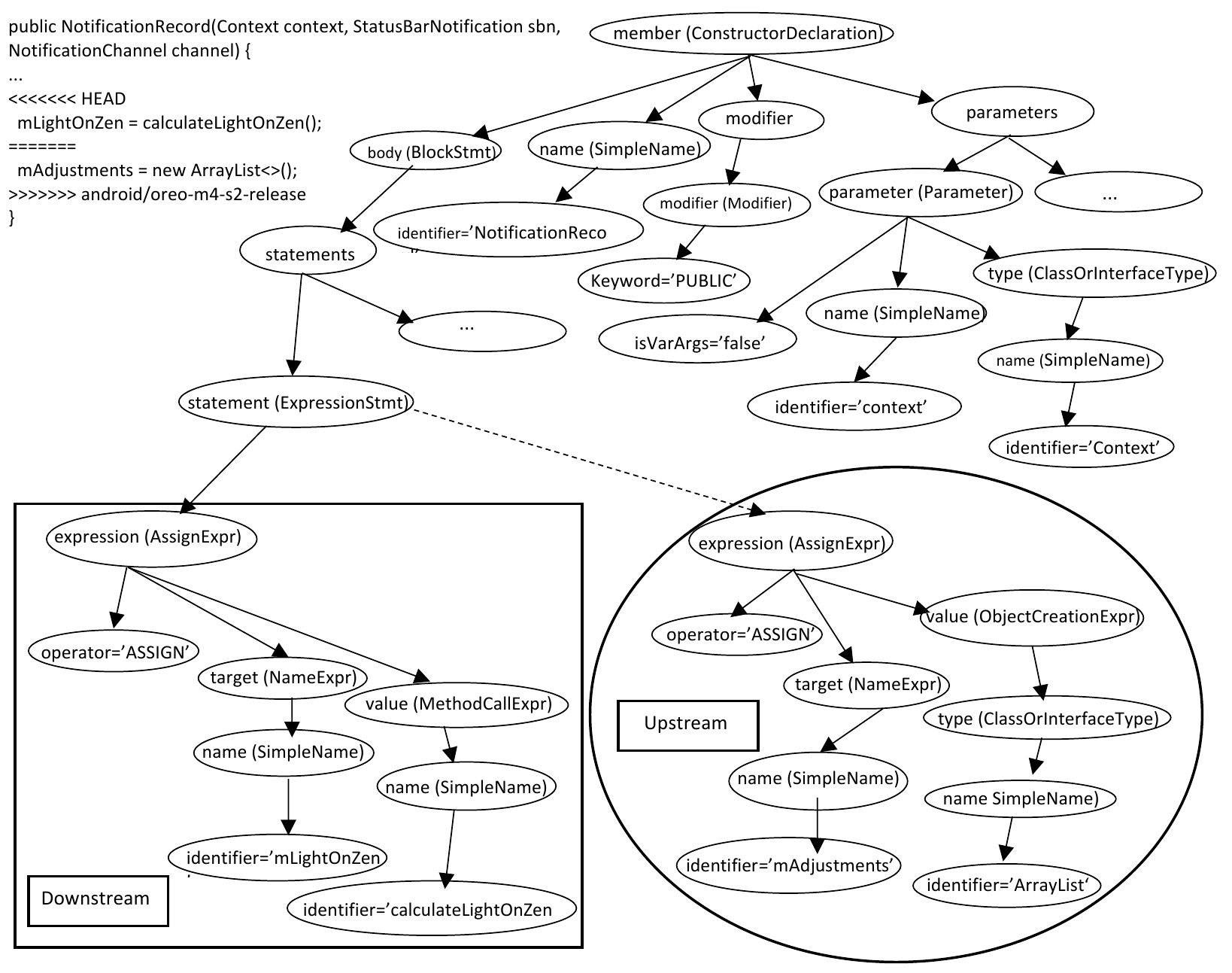}
    \vspace{-10pt}
    \caption{Sample AST.}
    \label{fig:method_ast}
    \vspace{-20pt}
\end{figure}

\begin{figure}[t!]
    \centering
    \begin{minipage}[b]{0.99\linewidth}
         \centering
         \includegraphics[width=\textwidth]{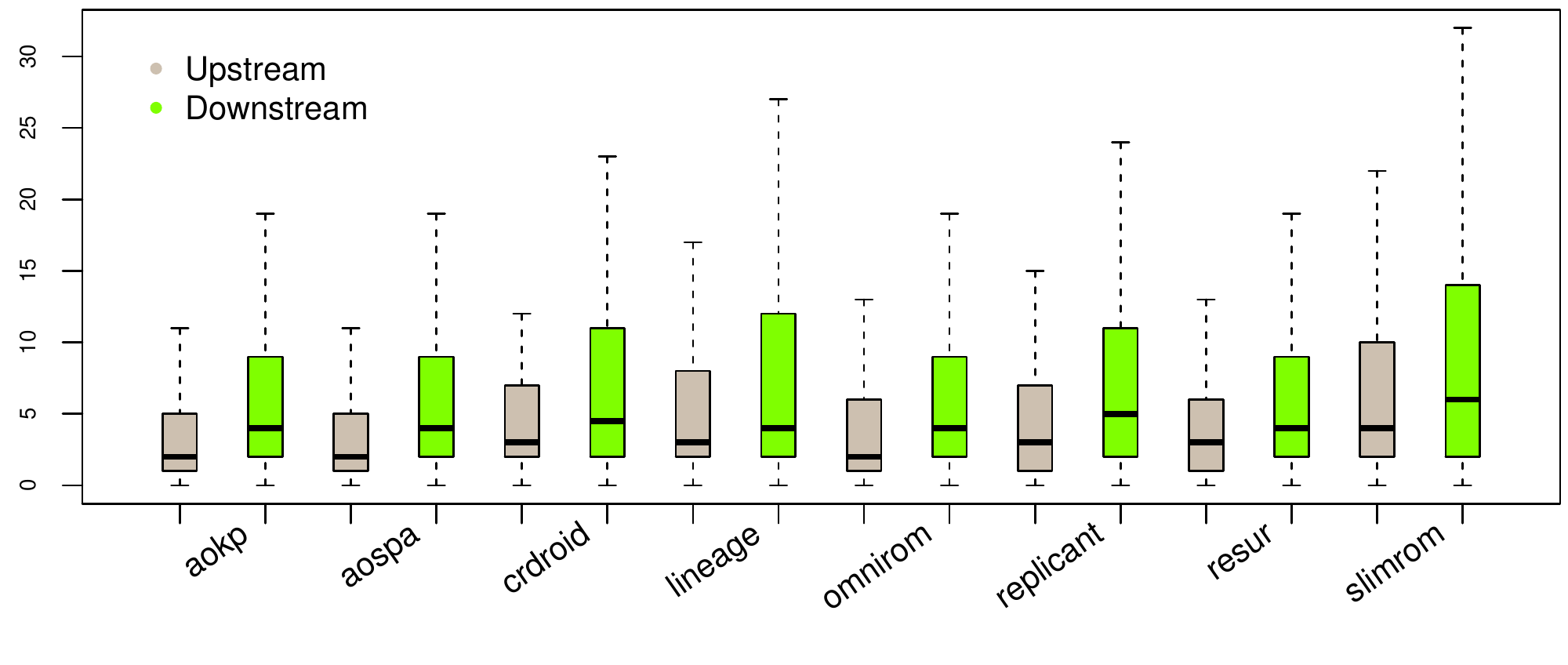}
         \vspace{-32pt}
         \caption{Lines affected by conflicts.}
         \label{fig:conflict_line}
    \end{minipage}
    \vspace{-15pt}
\end{figure}
\begin{figure}[t!]
    \begin{minipage}[b]{0.99\linewidth}
         \centering
         \includegraphics[width=\textwidth]{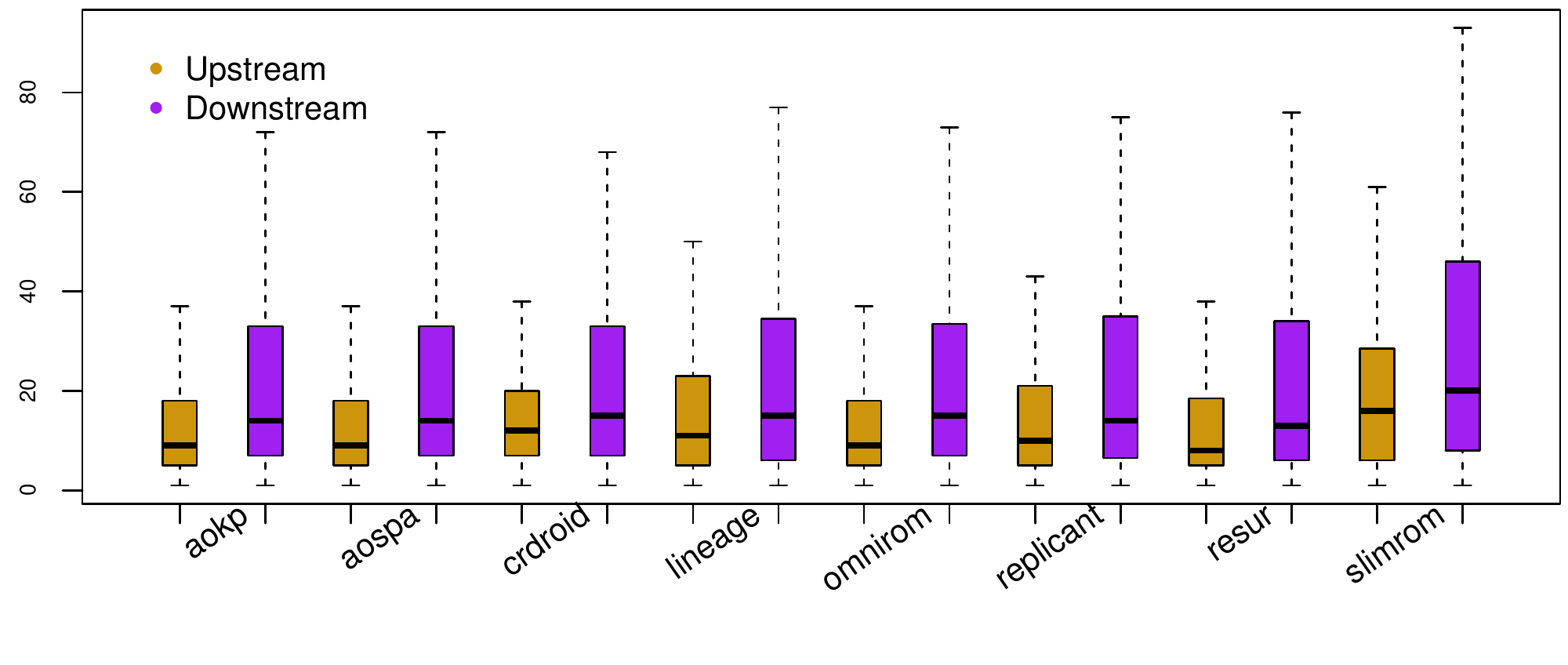}
         \vspace{-32pt}
         \caption{AST nodes affected by conflicts.}
         \label{fig:conflict_ast}
    \end{minipage}
    \vspace{-16pt}
    % \caption{The distribution of commits and tags.}
\end{figure}

Fig.~\ref{fig:method_ast} provides an example of our analysis operates. The example depicts a portion of the AST of the sample code shown in the top-left part of the figure. The square and round parts are the parts identified by our analysis as the ones affected by the conflict. The square part shows the AST nodes of the code in downstream project (line below {\small\texttt{$<<<<<<<$ HEAD}}) while the round part represents the code in the upstream repository (line below {\small\texttt{=======}}). In this RQ, we characterize the parts of the conflicts that are coming from both the downstream and upstream projects. We consider the parts from upstream because disregarding them might lead to compatibility issues.

\noindent\textbf{Results:} Fig.~\ref{fig:conflict_line} presents the distribution of the number of code lines involved in the code blocks affected by a conflict in both the upstream and downstream projects. \textbf{The number of code lines from the downstream project is always larger than that of the upstream project}. After running a Mann–Whitney–Wilcoxon (MWW) test, we confirmed that the difference is significant. As revealed in our manual investigation, the rationale behind this difference is that developers of the downstream project have attempted to regularly update some of the methods, and such changes are often more frequent than that of the upstream project. This evidence experimentally shows that large and possible complex changes are indeed involved in method-related conflicts.
Fig.~\ref{fig:conflict_ast} presents the distribution of the number of AST nodes involved in the conflicted code blocks in the upstream and downstream projects.
Similarly to results presented in Fig.~\ref{fig:conflict_line}, \textbf{the size of changed AST nodes in the downstream project is larger than that of the upstream project}. This characteristic is confirmed as significant by the MWW test. This result confirms our previous findings that the changes in method-related conflicts could be complex.

% \begin{figure}
%     \centering
%     \includegraphics[width=\linewidth]{pics/method_ast.pdf}
%     \caption{Conflicted asts during merge.}
%     \label{fig:conflict_ast}
% \end{figure}

\vspace{-10pt}
\begin{tcolorbox}[before skip=0.4cm, after skip=0.6cm, title=\textbf{RQ5 Findings}, left=2pt, right=2pt,top=2pt,bottom=2pt]
Conflicts affecting methods can involve a large number of AST nodes. Furthermore, the number of lines and AST nodes affected by conflicts are often higher in downstream projects than upstream projects. This result seems to indicate that developers significantly extend the functionality of upstream projects. When developers want to merge changes from upstream projects, they need to focus on a non-trivial portion of the code. 
\end{tcolorbox}

\subsection*{RQ6: \textit{How often downstream developers change files in their projects when they experience a merge conflict?}}
\label{subsec:rq6}

\noindent\textbf{Methodology:} With this research question, we are interested in understanding how often downstream developers change their projects when they experience a merge conflict. While answering this research question, we also look at how frequently developers disregard changes from upstream projects.
When resolving a conflict, developers have three choices: (i) integrate upstream and downstream code changes by combining code entities from the two merged branches (\textit{Integrate}), (ii) transform the downstream repository to have the same code as the upstream repository by removing the part of the conflict belonging to the downstream branch (\textit{Use Upstream}), and (iii) keep the downstream code by removing the part of the conflict belonging to the upstream branch (\textit{Use Downstream}). To answer the RQ, we classify conflict resolutions into one of the three categories by analyzing the ASTs of the downstream and upstream repositories before and after the merge operations.

\noindent\textbf{Result:} Table~\ref{tab:conf_ignore} summarizes the results of our analysis. The table reports the number of conflicts in the three categories we considered (\textit{Integrate}, \textit{Use Upstream}, and \textit{Use Downstream}). The table also summarizes the total number of code entities characterizing the conflicts in the last column. The considered customizations have a significant number of code entities in the \textit{Use Downstream} category. Specifically, the percentage of code entities in this category varies from 27.43\% to 89.26\%. This experimental result suggests that the \textbf{customizations of the Android framework stick to their changes during their evolution}. Although this characteristic is likely intended as customizations aim to offer different versions of the Android OS, the differences (especially if they involve semantic changes) will exacerbate the fragmentation problem characterizing the Android ecosystem and cause compatibility issues to users.
%mf: removed as we should talk about this in the discussion
%We believe that there is a need for developer-centric techniques to better explain the impact of ``ingnored'' changes to downstream downstream developers, as, ideally, such changes should not be ignored. This situation also calls for automated refactoring approaches for mitigating the potential compatibility issues caused by those changes. This is, however, out of the scope of this paper. We hence take it as our future work.

\vspace{-4pt}
\begin{tcolorbox}[before skip=0.4cm, after skip=0.6cm, title=\textbf{RQ6 Findings}, left=2pt, right=2pt,top=2pt,bottom=2pt]
Downstream developers are required to resolve a large number of conflicts. This situation calls for automated techniques to help developers in the task. Furthermore, our results also show that in more than 40\% of the cases, downstream developers disregard the portion of the conflict from the upstream repositories. This might lead developers to introduce compatibility issues.
\end{tcolorbox}

\begin{table}[t!]
    \centering
    \scriptsize
    \caption{Code entities in the different resolution categories.}
    \vspace{-5pt}
    \label{tab:conf_ignore}
    \resizebox{\linewidth}{!}{
    \begin{tabular}{r|c|c|c|c}
    \hline
        \textit{Downstream Project Name}            &   \textit{\makecell[c]{Integrate}}      &      \textit{\makecell[c]{Use\\ Upstream}}      &    \textit{\makecell[c]{Use\\ Downstream}}   &  \textit{Total} \\
    \hline
        \textsc{AOKP}               &        22                      &          13             &        43  (55.13\%)          &  78    \\
        \textsc{AOSPA}              &        809                     &          354            &        524 (31.06\%)          &  1,687 \\
        \textsc{crdroidandroid}     &        53                      &          21             &        119 (61.66\%)          &  193   \\
        \textsc{LineageOS}          &        375                     &          195            &        519 (47.66\%)          &  1,089 \\
        \textsc{OmniRom}            &        802                     &          439            &        469 (27.43\%)          &  1,710 \\
        \textsc{replicant}          &        77                      &          44             &        83  (40.69\%)          &  204   \\
        \textsc{ResurrectionRemix}  &        33                      &          15             &        399 (89.26\%)          &  447   \\
        \textsc{SlimRoms}           &        108                     &          50             &        235 (59.80\%)          &  393   \\
        
        % eos                &     0   (0\%)        &   0     \\
    \hline
        \textit{Total}  &  2,279  &  1,131 &  2,391 (41.22\%)  &  5,801 \\
    \hline
    \end{tabular}}
    %\$^\alpha$ The upstream changes are essentially ignored in these cases.
    \label{tab:conf_resolution}
    \vspace{-30pt}
\end{table}

\vspace{-5pt}
\subsection*{RQ7: \textit{Do developers of Android apps use methods affected by merge conflicts?}}  

\noindent\textbf{Methodology:} In this RQ, we are interested in investigating the use of conflict-affected methods in client Android apps. Since the implementation of the methods may be different between downstream projects and the official Android OS, there might be inconsistencies when customers are running apps containing these conflict-affected methods. The inconsistencies between these conflict-affected methods could lead to compatibility issues. There could be two kinds of compatibility issues. The first kind can appear when a downstream OS removes a method from the API that is instead present in the upstream OS. The second kind can appear when a method in the downstream OS has different semantics as compared to the same method in the upstream OS.

To empirically identify the extent to which conflict-affected methods are leveraged by real-world Android apps, we randomly select 1,000 apps from AndroZoo~\cite{li2017androzoo++}. The apps in AndroidZoo are collected from real-world app markets such as the official Google Play store.
For each of the selected app, we first disassemble it and scan its source code to collect all of its accessed methods from the Android API.
We then compute the intersection between the used methods from the Android API and the conflict-affected methods identified in this work.
If the result set is empty, we conclude that the app leverages no conflict-affected method.
Otherwise, we identify that the app leverages $x$ conflict-affected methods, where $x$ is the size of the result set.

\noindent\textbf{Results:} \textbf{644 (or 64.4\%) of 1,000 apps have accessed conflict-affected methods}. The fact that over half of the randomly selected apps access such methods demonstrates that the potential impact of customization changes could be huge.
Furthermore, analyzing the distribution of the number of accessed conflict-affected methods per app, we identified that \textbf{for most of the apps, there is more than one conflict-affected method accessed}, which further increases the possibility of encountering compatibility issues.
Based on these results, we argue that the downstream project developers need to pay special attention to resolving the conflicts when merging updates from their upstream projects.
Moreover, as a significant portion of conflicts has already been ignored by downstream developers (cf. the findings discovered in Section~\ref{subsec:rq6}), certain compatibility issues might have already been introduced to the field.
%We hence argue that there is also a need to devise automated approaches to mitigate such compatibility issues.

%mf: removed as it did not add anything relevant
% \begin{figure}
%     \centering
%     \includegraphics[width=\linewidth]{pics/app_conf_methods.pdf}
%     \caption{Conflict affected methods in 1,000 Android apps.}
%     \label{fig:conf_method_dis}
% \end{figure}

\begin{tcolorbox}[before skip=0.4cm, after skip=0.6cm, title=\textbf{RQ7 Findings}, left=2pt, right=2pt,top=2pt,bottom=2pt]
64.4\% of randomly selected Android apps have accessed methods affected by conflicts.
In more than half of the cases, the apps access more than one conflict-affected method.
The use of these methods could result in compatibility issues and approaches to detect these
potential issues are needed.
\end{tcolorbox}
%\item What are the merge conflict fix patterns?
%\item When merging the two systems, is it only the diffs are merged or the whole system?

\section{Discussion}
\label{sec:discussion}

%We now discuss the implications of our findings as well as the potential threats to the validity of this work.

%\subsection{Implications}

%The findings and observations of this study raise a number of issues and opportunities for researchers and practitioners.

%\begin{itemize}
%    \item (1) We found that there are a lot of conflicts. We need to build tools help automate updates from upstream projects.
\textbf{Continued research on techniques to assist updates from upstream is needed.} As most of the downstream projects that we considered encounter more than 40\% merge conflicts and resolving conflicts often involves considering a large number of code entities (RQ5), we argue that there is a strong need to continue devising automated approaches to help complete those merge-operations. This is especially true for helping customized Android frameworks keep the evolution pace with the official Android framework, which is one of the most actively maintained and largest open-source projects.
As argued by Owhadi-Kareshk et al., automated predictors could be also leveraged to predict potential merge conflicts, as a pre-filtering step for speculative merging~\cite{owhadi2019predicting}. Additionally, given the variance we identified in the merge practices across different projects, we believe that it is necessary to extend our results by performing studies that involve the developers of the customized projects to further understand constraints in their merge practices and even more suitably guide the creation of techniques and tools to merge from upstream.
    
\textbf{Merge changes from upstream as soon as possible.}
Downstream project developers should merge changes as soon as possible, since the less time lag before merging from upstream, the fewer merge conflicts they may encounter. Since downstream projects always need to be synced up with upstream to bring in features, bug fixes, etc., it is good to have routine merging approaches from upstream, which would relieve developers' burdens in resolving the merge conflicts.
It could be even better if automated approaches can be proposed to support such timely merge strategies.

\textbf{Automated techniques are needed to help developers handling comments conflicts.}
Downstream customizations have a large number of conflicts appearing in inline or stand-alone comments (e.g., comments in the middle of a class not directly associated with code). Since comments are of great importance for developers of these downstream projects, the conflicts associated with inline or stand-alone comments can be time-consuming to resolve as developers need to first relate the changes in the conflict to specific code sections. To improve the development of customized OSs, automated techniques are needed to relate comments changes to code changes. These techniques could leverage the information contained in the version control system and use static analyses combined to natural language processing to relate changes to code sections automatically.

\textbf{Merge details are needed to aid the evolution of the customized OSs.}
In the study, we identified that downstream developers do not follow a systematic approach for merging updates from upstream projects. Additionally, in our manual inspections, we identified that developers rarely use detailed commit messages to described the merge operations they performed. Because of this situation, different developers in the same project might not be aware of semantic changes, bug fixes, or security patches applied (and not applied) by other developers, and this might have negative consequences on their development activities. We suggest that developers of customized OSs follow merging strategies agreed upon in the projects or provide detail commit messages in their merge operations describing the changes applied to the project. To this end, automated approaches could suggests commit summaries based on the changes made in the merge operations. This advice also extends to files other than source code. For example, we identified a large number of xml files affected by merge conflicts, and those files can affect the UI or default configurations of the OS.

% Downstream project developers should merge official Android releases in an systematic approach that is periodically merge release tags from official Android releases, such as the very first release tag of the new official Android revision to demonstrate a new reversion of the downstream project and then merge periodically for example two to three weeks after last merge. In the meanwhile, merge other releases as needed as a complementary, such as security patches merge etc. Moreover, commit messages related to resolve merge conflicts induced by other files are also vital for developers of downstream, such as the rationale behind the update of png and xml files which is very important for UI changes since it would provide a better documentation both for downstream developers for next release and OS customers to know the difference with official Android release.
    %\item (2) We found that developers change certain methods after resolving conflicts. We need better automated testing techniques to know whether conflicts have been successfully resolved. We could devise techniques based on differential testing to identify weather the behavior with respect to the previous version of downstream is preserved and whether the new changes in upstream are successfully ported.

\textbf{Mitigated compatibility issues.}
As revealed in our experimental study, developers of downstream projects have ignored some of the changes from upstream projects.
%At the moment, it is still unknown what are the ignored changes responsible for.
Some of the changes, e.g., relevant to fixing bugs or augmenting features, may cause semantic differences between the two frameworks, which are supposed to support the execution of the same apps.
Subsequently, those changes could lead to potential compatibility issues, which are hard to identify and hard to account for app developers.
We argue that automated approaches are also needed to identify and mitigate instances of this situation, e.g., by (i) detecting semantic deviations from upstream projects, (ii) helping downstream project developers to better merge the changes of their upstream projects, or (iii) by helping app developers protect (or avoid) the usages of methods that may cause compatibility issues.
    %We found that developers ignore some of the changes. These changes could related to a bug fix. Mobile app developers do not know if their app is affected by behavior of a certain customization of the OS. We need technique for patch-presence testing and let developers know. These techniques should be tailored to the code in developers' apps.

%\end{itemize}

\section{Threats to validity}
%\textbf{Internal validity:}
%conflict resolve determine. Because we can not one hundred percent make sure if it has been ignored or not.
%\pei{Is False positive counted as internal validity?}

%\textbf{External validity:}
The primary \textbf{external threat to validity} of this study lies in the choice of downstream projects. In this study, we have only analyzed eight different open-source downstream projects and the results might not generalize to other projects. To mitigate this threat, we conducted experiments on all of the downstream projects that we could readily get and these projects include different customizations of the Android OS. We also believe that an interesting direction for future work could focus on extending our results by searching and considering customizations of the Android OS that are available on GitHub.
% Furthermore, the fact that all of the popular downstream projects have a similar evolutionary route, which is also in line with the evolution (based on their release messages) of popular closed-source downstream projects (i.e., MIUI~\cite{miuiwiki}, OnePlus~\cite{onepluswiki}), gives confidence in the fact that results might also generalize to other projects.
Furthermore, our research might not generalize to close-sourced projects such as MIUI~\cite{miuiwiki} or OnePlus~\cite{onepluswiki}. However, after inspecting the release messages of those projects, we noticed that those projects have update patterns that are in line with the ones we identified in our study. Although encouraging on the possibility that the results might generalize, additional studies based on those projects are needed to further understand upstream merge practices in the context of Android.
%Therefore, we believe that our finding can be extended to other downstream projects. 
%\textbf{Construct validity:} 
%In terms of construct validity, there might be errors in the implementation of the tools we built to perform the study. To mitigate this threat, we extensively inspected the results of the study manually.

The main \textbf{threat to construct validity} resides in the possibility that the implementation of our experimental scripts and tools to do the analysis might contain errors. To mitigate this threat, we extensively inspected the results of the study manually. For example, we determine if the conflict is handled by ignoring the code snippet from upstream projects through manually checking whether the code snippet exists in the Java files associated with the completed merge operation. The authors also cross-validated the results.

%\todo{any internal validity threats??}

\vspace{-5pt}
\section{Related work}
\label{sec:related}
%\mf{For each paper discussed in related work. We should have a short summary of what related work does and detail how our work is different from related work.}
There are other works that have studied merge conflicts~\cite{nguyen2015detecting, mens2002state, apel2011semistructured, mahmoudi2018android, sung2020towards, brun2011proactive, shao2009sca} in divergent projects/branches. In this section, we describe related work closely related to ours.

%Merge conflicts come from the parallel updates of the software by different artifacts developers who are not necessarily aware of each other's modifications. 
Mens~\cite{mens2002state} performed a survey to investigate a wide range of different merge techniques, from the initial pure textual merging to the more powerful approaches taking syntax and semantics into account. Various merge techniques can be classified into different categories from different perspectives. %, such as two-way or three-way merging, textual, syntactic, semantic or structural merging, and state-based, change-based or operation-based merging. 
In addition, the author in the paper compares these general merge techniques based on many different important criteria.%, such as expressiveness, accuracy, domain dependence, scalability, efficiency, and granularity. Additional investigation and research are needed to resolve the problem of merge conflicts. 
Different from the survey comparing several different merge techniques, we focus on the performed textual merge operations in different customized Android projects.

% Apel et al.~\cite{apel2011semistructured} discussed conflicts resolved in revision control systems. There exist, in general, two different revision control systems, including a structured revision control system that leverages language-specific knowledge for conflict resolution and an unstructured revision control system that utilizes textual similarity to resolve merge conflicts. To reap the advantages of both structured and unstructured revision systems to resolve conflicts, they proposed a semistructured revision system to gain the strengths of expressiveness of structured systems and generality of unstructured systems. %Their experiments on 24 different software artifacts written in Java, C\#, and Python consisting of 180 merge scenarios demonstrate that semistructured merge can reduce, on average, 34\% merge conflicts, compared to the unstructured text-based merge resolution. 
% Our work, however, focuses on the conflict details and the cost to resolve such conflicts, rather than different automatic merge techniques.

Mahmoudi et al.~\cite{mahmoudi2018android} carried out an empirical study to determine the details of the updates between Lineage OS % which were forked from the official Android, 
and the update of the official Android.
%When a new version of the official Android release, developers of customized Android OS who want to update their operating system indeed need to merge and re-apply their customized modifications to the new release.%, which is unavoidable, difficult, and time-consuming. 
%To perform their study, they selected the popular community-based variant of customized Android OS LineageOS as the proxy to perform the study. They attempted to analyze the overlap updates between LineageOS and the official Android OS in 8 different releases and proposed an open-source toolchain to facilitate the analysis. 
They found that 83\% of subsystems updated in LineageOS is also modified in official Android and that 56\% of the overlap modifications in LineageOS can be automatically safely merged into the next new release version. While they only focus on eight different update problems of the details of the changes applied in LineageOS versus those in Android, we highlight the merge conflicts during the evolution of Android but not limited to LinegeOS, which means they analyze the final result of the evolution but we highlight the merges during the whole evolution process.

Cavalcanti et al.~\cite{cavalcanti2017evaluating} and Accioly et al.~\cite{accioly2018understanding} carried out a comprehensive study on semistructured merge on the evolution of large-scale open-source Java projects. Accioly et al. conducted an empirical study on large-scale open-source Java projects to figure out the characteristics of merge conflicts during merge by a semistructured merge tool over 70,047 merges from 123 Github Java projects. %They found that 84.57 of merge conflicts happen due to editing the same lines, or consecutive lines of the same method and the number of conflicting merge scenarios can be reduced if the merge is operated in a advanced merge algorithm. In addition, 
They conclude that the merge conflicts usually involve more than two developers, which means they need to understand different branches developed by different developers to successfully merge.  Cavalcanti et al. did a comprehensive study over more than 30,000 merges from 50 open-source projects. 
%They reproduced the merge operations both in unstructured and semistructured manner and identified inconsistent conflicts merges between these two different merge approaches. 
They found that some of the conflicts induced by unstructured merge can be auto-removed and also the merge conflicts are easier to analyze and resolve. Therefore, they proposed an advanced semistructured merge tool combining both approaches when merging and the experiment results indicate the tool is promising in reducing the false positive merge conflicts. Different from these study, we did not involve semistructured merge but only focused on unstructured merge operation and their conflicts resolution.

Ghiotto et al.~\cite{ghiotto2018nature} performed an extensive comprehensive study on the merge conflicts in the histories of
2,731 open source Java projects. 
%They first manually analysed the merge histories of 5 projects and then automate the analysis steps on all 2,731 projects. 
They found that 40 percent of the failed merges have a single conflicting chunk and 90 percent have 10 or fewer. The majority of the conflicting chunks have up to 50 LOC (Line Of Code) in each version and the method invocation is the most frequent language construct. %To resolve the merge conflicts, they found that about one fourth of the merge conflicts are resolved by only using lines in two merge versions. 
Among different constructs in merge conflicts, they revealed that if statement, method invocation are of the most difficulties. % and patterns exist in both how certain kinds of conflicts are addressed repeatedly and how developers make similar resolution choices over time. 
We also did an extensive analysis of merge conflicts in the evolution of Android customizations. However, we only focus on the merge operations bringing in updates from upstream projects without taking branch or fix merges into consideration, which would have some negative effect on the synchronization merges from upstream projects.

Shen et al.~\cite{shen2019intellimerge} proposed a new merge approach called IntelliMerge. The approach provides a refactoring-aware merging algorithm for Java programs. The evaluation of the approach is based on 10 Java projects and  it reduces the number of merge conflicts as compared to related work chosen as the baseline. Nishimura et al.~\cite{nishimura2016supporting} also present a tool, MergeHelper, to help in merging independent development from different developers by replaying the detailed code changes related to the conflict class members.

Lamothe et al.~\cite{lamothe2021systematic} did an extensive systematic literature review of API evolution including Android and other Java-based APIs. Inspired by the approaches in existing literature~\cite{keele2007guidelines,petersen2015guidelines}, they utilized a systematic approach to initiate a survey related to API evolution. By collecting online literature from five well-known technical publishers, they summarized different challenges raised by different researchers and concluded with three dominant challenges, including API changes pinpointing, benchmark creation for the analysis of API evolution, and understanding the impact of API evolution.

\vspace{-5pt}
\section{Summary}
\label{sec:summary}

In this paper, we presented an empirical study that investigated how developers of customized versions of the Android OS update their projects to merge changes from the main version of the Android OS. In our study, we analyzed the merge operations from eight open-source customized Android OS projects and identified how these operations affect a randomly selected sample of 1,000 apps. Our results show that the developers performed a small percentage of the possible updates, the merge operations often lead to conflicts, and a large percentage of the apps considered use methods affected by conflicts. The large number of conflicts identified and the low percentage of merge operations performed motivate further research on automated or semi-automated techniques to support the merge operation task. Furthermore, the high number of conflicts and the high percentage of apps using methods affected by conflicts indicate that these apps might experience compatibility issues, motivating further research in helping app developers handle this type of issues.
As future research, we plan to present our findings to developers that work on customizations of the Android OS to determine changes they find most challenging to merge. We then plan to develop automated techniques to help developers correctly perform merge operations more frequently, efficiently and effectively.

\section*{Acknowledgements}

This work is supported by ARC Laureate Fellowship FL190100035, Discovery Early Career Researcher Award DE200100016, Discovery Project DP200100020.

% %The authors would like to thank the anonymous reviewers who have provided insightful and constructive comments to the conference version of this extension.
% This work was supported by the Australian Research Council (ARC) under a Laureate Fellowship project FL190100035, a Discovery Early Career Researcher Award (DECRA) project DE200100016, and a Discovery project DP200100020.

% \bibliographystyle{IEEEtran}
% \bibliography{IEEEabrv,main}

\bibliographystyle{ACM-Reference-Format}

\balance

\bibliography{main}

%\begin{IEEEbiography}[{\includegraphics[width=1in,height=1.25in,clip,keepaspectratio]{figures/Mattia}}]{Pei Liu} TODO.
%\end{IEEEbiography}

%\begin{IEEEbiography}[{\includegraphics[width=1in,height=1.25in,clip,keepaspectratio]{figures/Mattia}}]{Mattia Fazzini} is an Assistant Professor in the Department of Computer Science \& Engineering at the University of Minnesota. He completed his Ph.D. in Computer Science from the Georgia Institute of Technology in 2019. His research interests lie primarily in the area of software engineering, with emphasis on techniques for improving software quality. The central theme of his research is the development of approaches for testing and maintenance of mobile apps.  He has published in several top peer-reviewed software engineering venues including: ASE, ICSE, ICST, and ISSTA. He received the Facebook Testing and Verification award in 2019 for his research on mobile app testing. More information is available at \url{https://www-users.cs.umn.edu/~mfazzini/}.
%\end{IEEEbiography}

%\begin{IEEEbiography}[{\includegraphics[width=1in,height=1.25in,clip,keepaspectratio]{figures/Mattia}}]{Li Li} TODO.
%\end{IEEEbiography}

%\begin{IEEEbiography}[{\includegraphics[width=1in,height=1.25in,clip,keepaspectratio]{figures/Mattia}}]{John Grundy} TODO.
%\end{IEEEbiography}

\end{document}